\documentclass[aps,pre,twocolumn,superscriptaddress,longbibliography,floatfix]{revtex4-1}

\usepackage[utf8]{inputenc}
\usepackage[english]{babel}
\usepackage[T1]{fontenc}
\usepackage{lmodern}

\usepackage{amsmath,amsthm,amsfonts,amssymb,bm,amssymb}

\usepackage[colorlinks={true}, citecolor={blue}, filecolor={blue}, linkcolor={blue}, urlcolor={blue}]{hyperref}

\usepackage{graphicx}
\usepackage[caption=false]{subfig}

\usepackage{microtype}
\usepackage{soul}

\newcommand{\kB}{k_\mathrm{B}}
\newcommand{\ssF}{\scriptsize{F}}

\begin{document}

\title{Topological Phase Transition in Quantum Heat Engine Cycles}

\author{Mojde Fadaie}
\email{mfadaei@ku.edu.tr}
\affiliation{Department of Physics, Ko\c{c} University, 34450 Sariyer, Istanbul TURKEY}

\author{Elif Yunt}
\email{eyunt@ku.edu.tr}
\affiliation{Department of Physics, Ko\c{c} University, 34450 Sariyer, Istanbul TURKEY}

\author{\"{O}zg\"{u}r E. M\"{u}stecapl{\i}o\u{g}lu}
\email{omustecap@ku.edu.tr}
\affiliation{Department of Physics, Ko\c{c} University, 34450 Sariyer, Istanbul TURKEY}


\begin{abstract}
We explore signatures of a topological phase transition (TPT) in the work and efficiency of a quantum heat engine, which uses a single layer topological insulator, stanene, in an external electric field as a working substance. The magnitude of the electric field controls the trivial and topological insulator phases of the stanene.
The effect of TPT is investigated in two types of thermodynamic
cycles, with and without adiabatic stages. We examine a quantum Otto cycle for the adiabatic and
an idealized Stirling cycle for the non-adiabatic case. In both cycles, investigations
are done for high and low temperatures. It is found that Otto cycle can distinguish the critical point of the TPT as an extremum point in the work output with respect to applied fields at all temperatures. Stirling cycle can identify the critical point of the TPT as the maximum work point with respect to the applied fields only at relatively lower temperatures. As temperatures increase towards the room temperature maximum work point
of the Stirling cycle shifts away from the critical point of the TPT. In both cycles, increasing the temperature causes considerable enhancement in work and
efficiency from the order of meV to eV.

\end{abstract}

\maketitle
\section{Introduction}
Heat engines are the practical motivations and significant outcomes of the field of thermodynamics, which led to high socioeconomic impact in the industrial revolutions~\cite{fwt}. In parallel with the social and industrial developments, list of the working substances in the heat engines has been grown since the steam engine in 18th century, and included three-level masers~\cite{tlm}, cavity photons~\cite{ewf}, spin systems~\cite{dom}, single atom~\cite{asa}, atomic clusters ~\cite{sqh}, optomechanical systems ~\cite{qoh}, superconducting 
resonators~\cite{qhew}, Dirac particles~\cite{ooa,qhei}, graphene flake~\cite{mdq,mdqh,sgb}, black holes~\cite{hhe}, and ultracold atoms ~\cite{ath}. Heat engines that are using working substances requiring quantum mechanical descriptions are called as quantum heat engines (QHEs), whose cycles~\cite{qtca} are studied under a quantum thermodynamical framework, which is an emerging and rapidly progressing field of research~\cite{qtca,fao,itq,qt,tro}.  Experimental demonstrations of a single-atom heat engine~\cite{asa} and QHEs with nitrogen-vacancy centers in diamond~\cite{edo} and with cold Rb atoms~\cite{qheu} have been shown. 

Typically the working system of the engine remains in a single-phase during a thermodynamic cycle. In practice, however, phase changes can happen during the engine operation and can enhance the efficiency. Power plants, for example, can be modeled by the Rankine cycle, which is based upon water-steam phase transition. Recently it has been argued that diverging fluctuations at a second order phase transitions can help to achieve Carnot efficiency in a quantum Otto engine~\cite{tpo}. More recently, it has been shown that a quantum phase transition of an interacting spin working system allows for reaching Carnot efficiency~\cite{qtcw}. Inspired by these results, we ask if a topological phase transition (TPT)  during a quantum engine cycle can have any significant effects on the efficiency. Furthermore, we explore if such effects can be used to probe a TPT. In contrast to the ordinary QPTs, TPTs has no local
order parameter associated with symmetry breaking. TPT is described by a bulk invariant (Chern number) which is an integer and changes to another
integer at the TPT. In case of gap closing at the TPT, we expect that global nature of the work could capture the TPT qualitatively in certain cycles.

Usually, statistical work distribution measurements are studied in suddenly and infinitesimally quenched systems to examine phase transitions in quantum critical models~\cite{waq}.  Quench protocols bring the systems out of equilibrium and the work output is given by a probability distribution whose characteristic function can be related to the Loschmidt echo that can be determined in principle experimentally. It is shown that local quenches lead to edge singularities in the work distribution in a quantum critical system~\cite{sot}. Our approach is on the other hand based upon a cyclic variation of a control parameter, instead of quenching, and to look for signatures of TPT in the work output of the cycle. 

We emphasize that probing the TPT by the work an efficiency of QHEs is possible for those 2D materials, such as stanene, germanane and silicene, for which it is established that bulk gap closing (under electric field variation for example) is associated with a topological phase transition~\cite{murakami2011gap,ezawa2012topological}. We specifically consider a 2D monolayer Stanene (Sn) as our working substance. Stanene is a counterpart of graphene for tin atoms~\cite{lgq,lgqs,ezawa2012topological} with low-buckled honeycomb geometry ~\cite{tao}. In contrast to graphene it has a larger spin-orbit coupling~\cite{lee,fadaie2017investigation}, can host the quantum spin Hall effect at room temperature for dissipationless electric currents~\cite{lgq}, and its band structure, in particular, Dirac cone band gap, can be controlled with an out-of-plane electric field such that it exhibits a TPT between two-dimensional trivial and topological insulator (2DTI) phases depending on the applied electric field ~\cite{ttt,fadaie2016effect,murakami2011gap,moore2007topological}. 2D Sn has been fabricated recently by molecular beam epitaxy~\cite{ego}. Studies of thermal properties of Sn is limited to thermoelectrics so far~\cite{etp}. Its outstanding properties, on the other hand, make it also an ideal candidate for a working substance of a room temperature topological QHE. We note that using heat, instead of work, has been considered to detect dynamical phase transitions and Majorana modes recently~\cite{fmf}. It is proposed there that Floquet-Majorona phases can be used for QHEs or heat pumps. There are also theoretical investigations of
TPTs from the perspectives of Hill thermodynamics ~\cite{morais1,morais2}
and Uhlmann phase ~\cite{delgado1,delgado2}. Our simple model here can be envisioned as the first step towards QHEs with more sophisticated topological materials with TPTs. 

By using an external electric field as the control parameter we consider adiabatic (specifically the Otto cycle) and non-adiabatic (specifically a Stirling type cycle as in Ref.~\cite{qtcw}) thermodynamic cycles in which Sn undergoes a topological phase transition. We calculate the work output and efficiency of the cycles; determine and compare the signatures of the TPT in both types of cycles. We find out that the answer to our question for both cycles is positive, TPT of Sn can be probed by using work and efficiency of QHE cycles. Highly distinct characteristic behaviors are obtained below and above the critical point of TPT. Another
advantage of Sn working substance is that the topological QHE can operate around room temperature without the need for large external electric fields. We briefly discuss how to implement the cycle in an experimental setting a graphene bilayer as a scaffold at the end of the manuscript. 

This paper is organized as follows: We review the theory and TPT of Sn in Sec.~\ref{sec:workSys}. 
The results and discussions are presented in Sec.~\ref{sec:res} in two subsections where
quantum Otto cycle case is presented in Sec.~\ref{sec:Otto} and non-adiabatic cycle case is
presented in Sec.~\ref{sec:nonAdiabat}. In Sec.~\ref{sec:experiment}, a possible experimental realization is discussed. We
conclude in Sec.~\ref{sec:conc}. The details of the calculations of the work output for both of the quantum thermodynamic cycles are given in the Appendices~\ref{app:A} and~\ref{app:B}.
 \section{Working Substance}\label{sec:workSys}
This section is a brief review of relevant properties of Sn for our QHE and TPT discussions.  
We consider a single layer of two-dimensional Sn in the $xy-$plane as the working substance for a QHE. An external electric field $\varepsilon_z$ is applied in the  $z-$direction, perpendicular to the atomic layer. 
The system is described by a second-nearest-neighbor tight-binding 
model given in Ref~\cite{ezawa2012topological}
\begin{eqnarray}\label{eq:hamil}
 H&=&-t\sum_{<i,j>\alpha}c^\dagger_{i\alpha}c_{j\alpha}
+i\frac{\lambda_{\text{SO}}}{3\sqrt{3}}
\sum_{\ll i,j\gg\alpha\beta}
\upsilon_{\text{ij}}c^\dagger_{{i\alpha}}\sigma^z_{{\alpha\beta}}c_{j\beta} \nonumber\\
&-&i\frac{2\lambda_\text{R}}{3}
\sum_{\ll i,j\gg \alpha\beta}\mu_{\text{ij}}c^\dagger_{i\alpha}
(\vec{\sigma}\times{\vec{d^0_{ij}}})^z_{\alpha\beta}  c_{j\beta}\nonumber\\
&+&l{\sum_{i\alpha}}\zeta_i\varepsilon^i_zc{^\dagger_{{i\alpha}}c_{i\alpha}}.
\end{eqnarray}
The first term is the nearest-neighbor hopping term and $t$ is the transfer energy. The sum is taken over all pairs $<i, j>$ of the nearest neighboring sites, $c{^\dagger_{{i\alpha}}}$ and $c_{j\alpha}$ create and annihilate an electron with spin polarization $\alpha$ at site $i$, respectively. The second and third terms are the effective and intrinsic Rashba spin-orbit interactions with the corresponding coefficients $\lambda_{\text{SO}}$ and $\lambda_\text{R}$, respectively. $\mu_{\text{ij}}$ and $\zeta_i$ are equal to $\pm{1}$ for two sublatices of Sn.  $\vec{\sigma}$ denote the Pauli matrices; The coefficients 
$\upsilon_{\text{ij}}$ are defined as
\begin{eqnarray}
\upsilon_{\text{ij}}= {\vec{d_i}\times{\vec{d_j}}/ |\vec{d_i}\times{\vec{d_j}}|}
\end{eqnarray}
 where $\vec{d_i}$ and $\vec{d_i}$ are the two nearest bonds connecting the next-nearest neighbors $\vec{d}_{ij}$
and $ \vec{d}^0_{ij}=\vec{d}_{ij}/ |\vec{d}_{ij}|$. $l$ is half of the perpendicular distance between two sublattices. The sum is taken over all pairs $<<i, j>>$ of the next-nearest neighbors. We note that the model describes other two-dimensianal honeycomb structure as well~\cite{ezawa2012topological,ezawa_monolayer_2015}. 

 The low-energy effective Hamiltonian for stanene is derived from Eq.~(\ref{eq:hamil}) around the $K_\eta$ point in 
 Ref.~\cite{lee,ezawa2012topological} as  
 \begin{equation}\label{eq:ham1}
 H^s_{\eta}={\hbar v_f({\tau_x}k_x-\eta k_y\tau_y )}-\eta  \lambda_{\text{SO}}\tau_z s_z+l\varepsilon_z\tau_z,
\end{equation}
 where $\eta=\pm$1 is for the $K$ and $K'$ point, $v_f$ is the Fermi velocity and $\tau_{x,y,z}$ are the Pauli matrices of the sublattice. $s_z$ is the third Pauli matrix depicting the spin in the $z-$ direction.
 Here, we ignore the Rashba term in Ref.~\cite{ezawa2012topological}, as we will
consider the behavior of the bands exactly at the at the Dirac points where the Rashba term vanishes.
 The energy spectrum of Eq.~(\ref{eq:ham1}) is found to be~\cite{ezawa2012topological}
 \begin{equation}\label{eq:energyseta}
 E^s_{\eta}(k)=\pm\sqrt{\hbar^2v_f^2k^2+(l\varepsilon_z-\eta \lambda_{\text{SO}})^2},
 \end{equation}
 where $k=\sqrt{k_x^2+k_y^2}$. There are in total four distinct energy eigenvalues, 
 which are each two-fold degenerate.

\begin{figure}[!htp]
\includegraphics[width=8.3cm]{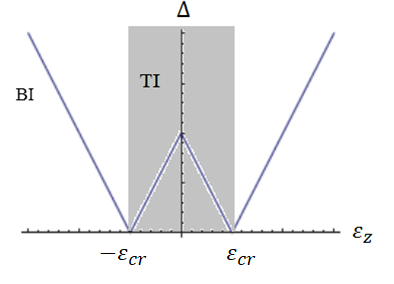}
\centering
\caption{(Color Online) Variation of band gap $\Delta$ as a function of external electric field 
$\varepsilon_z$. The gap is closed at the critical points 
$\varepsilon_z= \pm\varepsilon_{\text{cr}}$. Band closing is associated with a topological phase change for Sn
such that for $|\varepsilon_z|\textgreater\varepsilon_{\text{cr}}$ Sn is a band insulator (BI): while 
otherwise
it is a topological insulator (TI).}
\label{fig:TPT2}
\end{figure}

Low energy band structure of Sn at the vicinity of $K$ point given by Eq.~(\ref{eq:energyseta})
shows that the energy gap $\Delta=2|\lambda_{\text{SO}}-l\varepsilon_z|$ at $k=0$ is finite at $\varepsilon_z=0$; it 
decreases and closes at critical value $\varepsilon_{cr}$.
The phase diagram of Sn with respect to the external electric field is shown in Fig.~\ref{fig:TPT2}. 
The gap closing occurs at $\varepsilon_z= \pm\varepsilon_{cr}$ . 
It is found that gap closing is associated with TPT such that for $|\varepsilon_z|\textless\varepsilon_{\text{cr}}$ Sn is a TI and for $|\varepsilon_z|\textgreater\varepsilon_{\text{cr}}$ it is a band insulator~\cite{murakami2011gap,moore2007topological,ezawa2012topological}.  

\section{Results and Discussion}\label{sec:res}
\subsection{Quantum Otto Cycle} \label{sec:Otto}
\begin{figure}[htp!]
\includegraphics[width=8.3cm]{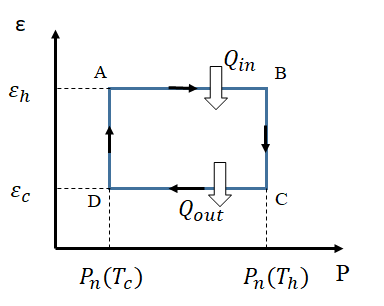}
\centering
\caption{Quantum Otto cycle operating between a hot bath at temperature $T_h$ and a cold bath at temperature $T_c$. It has of two isochoric (A to B and C to D) and adiabatic (B to C and D to A) processes for Sn under electric field $\varepsilon_z$. $Q_{\text{in}}$ is the heat injected in the A-B stage, and  $Q_{\text{out}}$ is the heat ejected in the C-D stage. Electric field changes between $\varepsilon_h$ and $\varepsilon_c$. Occupation probabilities $P$ of an energy level $E_n$
changes between $P_n(T_c)$ and $P_n(T_h)$.}
\label{fig:qoc}
\end{figure}

A quantum Otto cycle (QOC) consists of two isochoric and two adiabatic processes~\cite{qtca}. We consider a QOC with topological insulator as a working substance. The direction of the cycle is chosen such that net positive work is produced, as is depicted in Fig.~\ref{fig:qoc} . The  four stages of the QOC are as follows:
\begin{itemize}
	\item 	Stage 1 (A to B): This is a quantum isochoric process where the working substance in an external electric field, $\varepsilon_h$, with energy levels $E_n^h$ is coupled to a hot bath at temperature $T_{h}$. At point B, the working substance reaches thermal equilibrium with the hot bath, and the occupation probability of each eigenstate becomes $P_n(T_h)$, while the energy levels remain the same. No work is done but heat $Q_{\text{in}}$
\begin{equation}\label{eq:qin}
Q_{\text{in}}= {\int \frac{d^2\vec k}{(2\pi)^2}\sum_{n}g(E_n)E_n^h(k)[P_n(T_h)-P_n(T_c)]}
\end{equation}
 is absorbed by the working substance during this process. The factor $g(E_n)$ gives the degeneracy of $n^{\text{th}}$  energy level. The integral over $k=|\vec k|$ is from 0 to infinity. The
low-energy spectrum however is valid up to a large $k\sim 1/a$, where $a$ is the lattice constant of Sn.
The temperatures we consider are low enough to make sure that the large $k$ values cannot contribute to the
integral and hence we can use the low-energy spectrum instead of the full tight-binding solutions. 

	\item 	Stage 2 (B to C): In this process, which is a quantum adiabatic process, the working substance is isolated from the heat bath and the electric field changes from $\varepsilon_h$ to $\varepsilon_c$, where 
$\varepsilon_c<\varepsilon_h$. The energy levels change from $E_n^h$ to $E_n^c$. The occupation probabilities do not change. Work is done but no heat is transferred.
	\item 	Stage 3 (C to D): The working substance is subject to a constant electric field, $\varepsilon_c$, and is coupled to a cold bath at temperature $T_c<T_h$. The occupation probabilities at the end of this stage are $P_n(T_c)$. Heat $Q_{\text{out}}$
\begin{equation}\label{eq:qout}
Q_{\text{out}}={ \int \frac{d^2\vec k}{(2\pi)^2}\sum_{n}g(E_n)E_n^c(k)[P_n(T_c)-P_n(T_h)]}
\end{equation}
is ejected from the system.
In  Eq.~\ref{eq:qin} and Eq.~\ref{eq:qout}, $P_n(T_i)=f_n(k,\varepsilon_i,T_i)$  with $i=h,c$,  $\beta_{i}=1/\kB T_i$  where $\kB$ is the Boltzmann constant. $f_n(k,\varepsilon_i,T_i)=1/(\exp{[\beta_i(E^i_n(k)-E_{\ssF})]}+1)$ is the Fermi distribution function. $E_{\ssF}$ is the Fermi energy.

	\item Stage 4 (D to A): The system is separated from the cold bath and undergoes another quantum adiabatic process, as the electric field is changed from $\varepsilon_c$ to $\varepsilon_h$. Energy levels change from $E_n^c$ to $E_n^h$ The occupation probabilities remain the same. There is no heat transfer but work is done.
\end{itemize}

As the probability distribution remains invariant in the two quantum adiabatic processes,  the entropy will remain invariant as well and no net heat is produced. Based on this fact, the net work produced during a QOE cycle is given by
\begin{eqnarray}\label{eq:work}
W_{\text{O}}=&& \ Q_{\text{in}}+Q_{\text{out}}\nonumber\\
=&& \int \frac{d^2 \vec k}{(2\pi)^2}\times\nonumber\\
&&\sum_{n}g(E_n)\big(E_n^h(k)-E_n^c(k)\big) \big[P_n(T_h)-P_n(T_c)\big]
\end{eqnarray}

The efficiency of the heat engine is  $\eta_{\text{O}} = W_{\text{O}}/{Q_{\text{in}}}$. Positive work condition requires 
$Q_{\text{in}}> -Q_{\text{out}}$, under which the system would opearate as a heat engine. 
 

 There are a number of discussion in the literature on the thermodynamics of graphene type materials with positive and negative energy bands. It is the common treatment to limit the evaluation of the partition function to the positive energy manifold and ignore the contribution of the negative energies which is divergent~\cite{noncomm,diracoscthermal,boumali,eigenspectra}. Alternatively, a physical argument is suggested to eliminate the negative energies from partition function calculation by assuming doped graphene~\cite{mdq,weaklymodgraphene}. Here we take into account both the positive and negative energy bands in both physically and mathematically rigorous footing for a neutral Sn. The method is presented in Appendices.

We take typical values for the parameters of Sn TI model~\cite{lee} where 
$\lambda_\text{SO}=30$ meV. We let $l\varepsilon_{h}$ and $l\varepsilon_{c}$ change in the ranges of $0-40$ meV and $0-150$ meV for low and high temperatures, respectively. We separately consider 
high and low temperature operations of the engine. For the high temperature case we take the temperatures of the hot and the cold baths as  $T_h=300$ K and $T_c=150$ K, respectively; while for the low temperature case we assume $T_h=40$ K and $T_c=30$ K.

\begin{figure}[!htp]
	\centering
	\subfloat[]{\includegraphics[width=8.3cm]{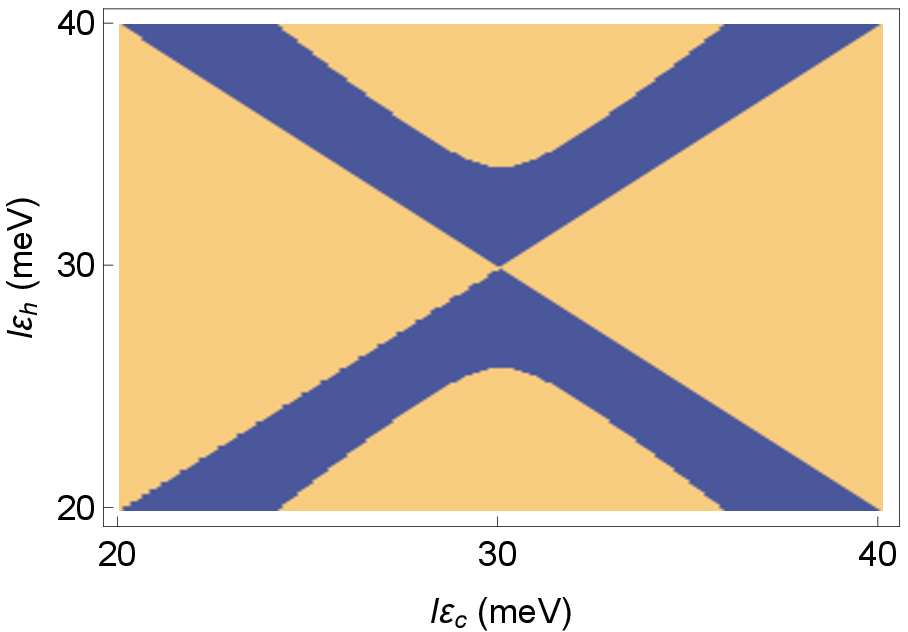}\label{fig:pwc3040}}\\
	\subfloat[]{\includegraphics[width=8.3cm]{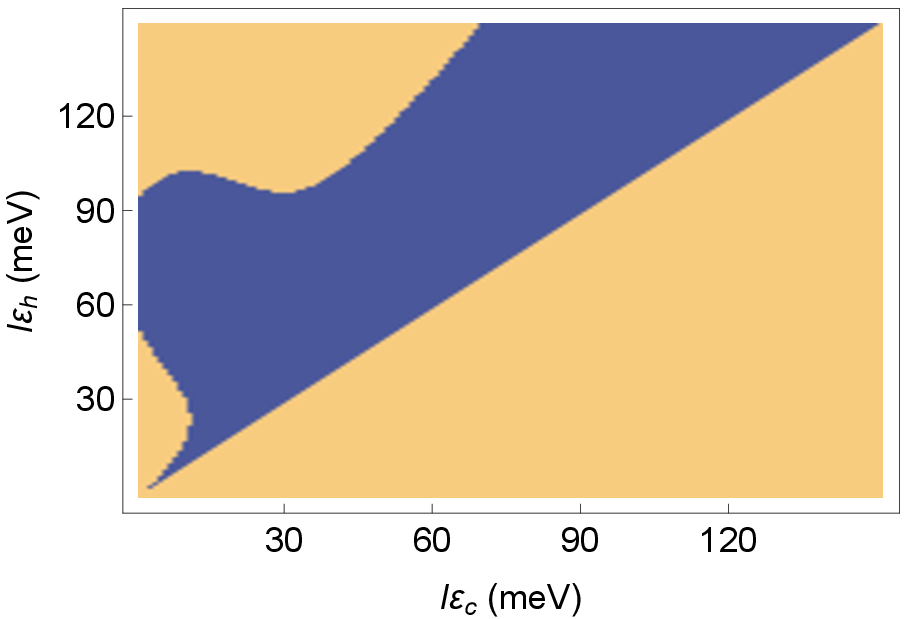} \label{fig:pwc300150}}
	\caption{{(Color Online)Domain of positive work in the space of external electric potentials at the cold and hot isochores $l\varepsilon_{c}$ and $l\varepsilon_{h}$, respectively. 
			Dark blue and light orange shaded regions indicate the domains of the positive and negative work.
			The parameters are the spin-orbit coupling $\lambda_{\text{SO}}=30$ meV, 
			and the temperatures of the hot and the cold baths which are taken to be
 (a)  $T_h=40$ K and $T_{c}=30$ K and 
(b) $T_h=300$ K and $T_{c}=150$ K, respectively.}}
	\label{fig:pwc}
\end{figure}

In Fig.~\ref{fig:pwc}, we show the electric potential domain of the positive work, $W_{\text{O}}>0$,
as the dark (blue) shaded region: while the negative work domain is indicated as light orange
shaded region. The positive work region for low temperatures plotted in Fig.~\ref{fig:pwc3040} shows an X-shaped structure, which deforms as the temperature increases. 
The positive work region for high temperatures shown in Fig.~\ref{fig:pwc300150} . It expands with the increasing electric fields, except for
a semicircular region terminated by the critical points of TPT at $l\varepsilon_{\text{cr}}=\lambda_{\text{SO}}$. 
Fig.~\ref{fig:pwc300150} could also be plotted for the negative values of the electric fields. The result would be 
inversion of the positive work regions in Figs.~\ref{fig:pwc3040} and~\ref{fig:pwc300150} with respect to the origin. We note that it is possible to extract positive work from the system even for $l\varepsilon_{h}<l\varepsilon_{c}$ for the low temperature regime according to Fig.~\ref{fig:pwc3040}.


\begin{figure}[!htp]
	\includegraphics[width=8.3cm]{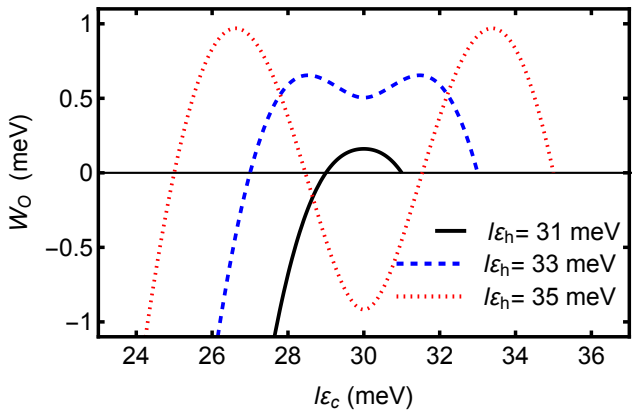}
	\caption{(Color Online) Work output, $W_\text{O}$ of the quantum Otto cycle operating between the hot and cold bath temperatures $T_h=40$ K and $T_{c}=30$ K, respectively, as a function of the electric potential at the cold isochore $l\varepsilon_{c}$ for different values of the electric potential
	at the hot isochore $l\varepsilon_{h}=31$ meV (black solid), $l\varepsilon_{h}=33$ meV (blue long dashed) and $l\varepsilon_{h}=35$ meV  (red short dashed). SO coupling is $\lambda_{\text{SO}}=30$ meV.}
	\label{fig:work3040}
\end{figure}

\begin{figure}[!htp]
	\includegraphics[width=8.3cm]{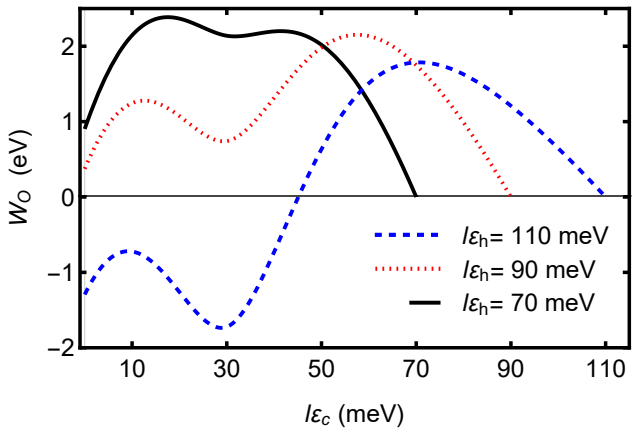}
	\caption{(Color Online) Work output, $W_\text{O}$ of the quantum Otto cycle operating between the hot and cold bath temperatures $T_h=300$ K and $T_{c}=150$ K, respectively, as a function of the electric potential at the cold isochore $l\varepsilon_{c}$ for different values of the electric potential
	at the hot isochore $l\varepsilon_{h}=110$ meV  (blue long dashed), $l\varepsilon_{h}=90$ meV (red short dashed), and $l\varepsilon_{h}=70$ meV (black solid). SO coupling is $\lambda_{\text{SO}}=30$ meV.}
	\label{fig:work300150}
\end{figure}

Exact values and behavior of the work output as a function of $l\varepsilon_{c}$, for three representative values of $l\varepsilon_{h}$ are plotted in Fig.~\ref{fig:work3040} for low temperatures, and in Fig.~\ref{fig:work300150} for high temperatures.
  Fig.~\ref{fig:work3040} shows a double peak profile. Zeros of the work function separate the regimes of the heat engine
and heat pump or refrigerator operations of the system. The number of zeros can be deduced from Fig.~\ref{fig:pwc3040} by the boundary between the positive and negative work domains. 
The critical point of TPT reveals itself as an extremum point in the curves at $l\varepsilon_{c}=\lambda_\text{SO}=30$ meV. Controlling $l\varepsilon_{h}$ can change the operation of the cycle from refrigerator to a heat engine.

 For high temperatures, the same double peak structure remains but with less symmetry between the topological and trvial insulator phases. The work value is enhanced by three orders of magnitude as shown in Fig.~\ref{fig:work300150} compared to low temperatures. It is worth noting that in the high temperature regime greater electric potential values are required to operate in the heat engine regime and, also, to observe the topological phase transition structure. 

\begin{figure}[!htp]
	\includegraphics[width=8.3cm]{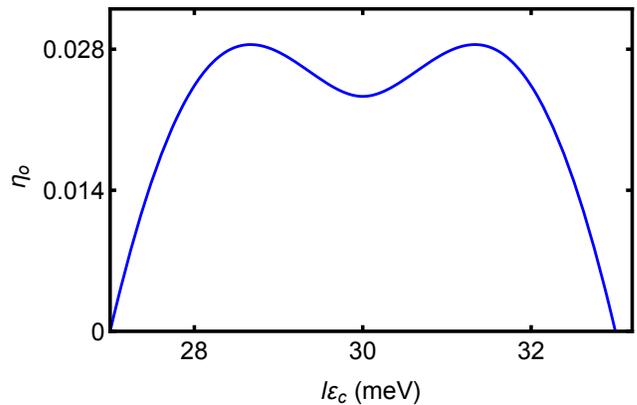}
	\caption{{(Color Online) Efficiency $\eta_{\text{O}}$ of quantum Otto cycle at low temperatures for the electric potential at the hot isochore $l\varepsilon_{h}=33$ meV, spin-orbit coupling $\lambda_{\text{SO}}=30$ meV,  and the temperatures of the hot and the cold baths $T_h=40$ K and $T_{c}=30$ K, respectively.}}
	\label{fig:efflow}
\end{figure}

\begin{figure}[!htp]
	\includegraphics[scale=0.9]{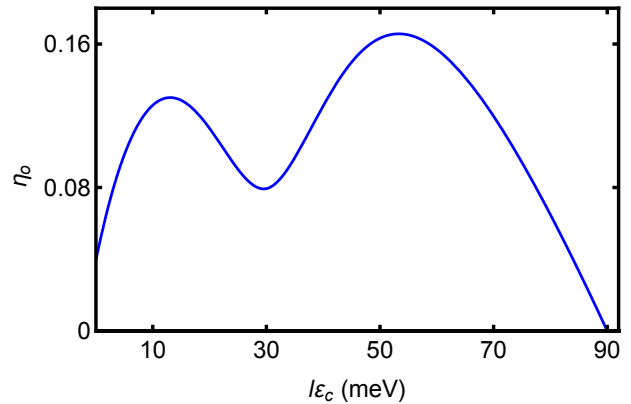}
	\caption{{(Color Online) Efficiency $\eta_{\text{O}}$ of quantum Otto heat engine at high temperatures for the electric potential at the hot isochore $l\varepsilon_{h}=90$ meV, spin-orbit coupling $\lambda_{\text{SO}}=30$ meV,  and the temperatures of the hot and the cold baths $T_h=300$ K and $T_{c}=150$ K, respectively.}}
	\label{fig:effhigh}
\end{figure}

The double peak profile appears in the efficiency behavior as well, as shown in Fig.~\ref{fig:efflow}, where we take $l\varepsilon_{h}=35$ meV, for low-temperature.   Qualitatively
similar behaviors are found for the high-temperature case. Fig.~\ref{fig:effhigh} indicates that the efficiency is higher for the high-temperature case, where $l\varepsilon_{h}=90$ meV.

  We remark that unless special procedures such as transitionless drives are used~\cite{transitionless}, 
speed of the cycle would be limited by the quantum adiabatic theorem. 
An alternative is to remove the adiabatic stages from the cycle completely. In the following subsection, we will consider a non-adiabatic cycle, which is used in Ref.~\cite{qtcw} to explore signatures of a quantum phase transition.

\subsection{Stirling Cycle }\label{sec:nonAdiabat}
\begin{figure}[htp!]
	\includegraphics[width=8.0cm]{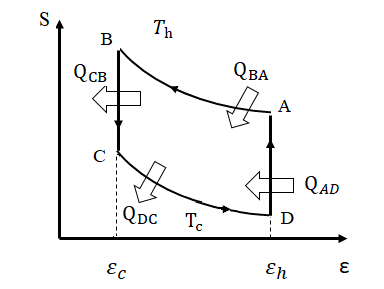}
	\caption{{Entropy-Electric field graph of a Stirling cycle consisting of two isothermal and two isoelectric stages. During the isothermal process from A to B (C to D), the system is brought into contact with a heat bath at temperature $T_h$  ($T_{c}$) and $Q_{BA}$ ($Q_{DC}$) amount of heat absorbed (released). In the isoelectric processes, the external electric field decreases (increases) from $ \varepsilon_h$ ($\varepsilon_c$) to $\varepsilon_c$ ($\varepsilon_h$) and the  exchanged heat is $Q_{CB}$ ($Q_{AD}$).}}
	\label{fig:cycle2}
\end{figure}

It is reported in Ref.~\cite{qtcw} that signatures of a quantum phase transition, associated with a level crossing, 
can be found in the work and efficiency of a thermodynamic cycle, which consists of two isothermal and two isomagnetic processes. Isothermal processes
transform the system through the phase transition point. TPT
of Sn is associated with the band closing and hence we may expect similar signatures of TPT in a similar cycle considered in Ref.~\cite{qtcw}.
The cycle can be compared to that of an idealized Stirling cycle without regenerator. We will call 
the cycle in Ref.~\cite{qtcw} as Stirling cycle.
In the case of Sn the isomagnetic stages are replaced by the isoelectric processes as shown in Fig.~\ref{fig:cycle2}.
However, the energy level structure of Sn is different than the model (interacting spins) in Ref.~\cite{qtcw}. At the K point, Sn can be considered as a four-level system. The only levels crossing are the two middle ones (lowest conduction and highest valence levels). 

The heat exchanged between the system and its surroundings in each stage of the cycle are given by
\begin{eqnarray}
Q_{BA}&=&T_h(S(B)-S(A)),\; Q_{CB}=U(C)-U(B), \label{qinsuncycle} \\
Q_{DC}&=&T_c(S(D)-S(C)),\; Q_{AD}=U(A)-U(D), \label{qoutsuncycle}
\end{eqnarray}
where $U(I)$
is the internal energy at constant electric field at each point $I=A,B,C,D$ in the cycle and $S(I)$
is the respective entropy. 
The total work produced in this cycle is given by 
\begin{eqnarray}
W_{\text{S}}=Q_{AB}+Q_{BC}+Q_{CD}+Q_{DA}.
\label{ws1}
\end{eqnarray}
Plugging  Eqs.~(\ref{qinsuncycle}) and ~(\ref{qoutsuncycle}), the total work is obtained in terms of the grand canonical partition functions ${\cal {Z}}(I)$ :
\begin{eqnarray}
W_{\text{S}}=\int\frac{d^2\vec k}{(2\pi)^2}&&\Big(\frac{1}{\beta_h}\ln\big({\cal {Z}}(B)\big)-\frac{1}{\beta_h}\ln\big({\cal {Z}}(A)\big)\nonumber\\ &&+\frac{1}{\beta_c}\ln\big({\cal {Z}}(D)\big)-\frac{1}{\beta_c}\ln\big({\cal {Z}}(C)\big)\Big)
\label{wstirling}
\end{eqnarray}	
where
\begin{equation}
{\cal {Z}}(I)=\prod_{n}[1+e^{-\beta E^i_n(k)}]^2.
\end{equation}
The energy of the system $E^i_{n}(k)$ at point $I$ is given by Eq.~(\ref{eq:energyseta}). The definitions of internal energy and entropy and also the details of the calculations leading to Eq.~(\ref{wstirling}) are given in Appendix~\ref{app:B}.

We investigate high and low temperature operation of the engine separately.
For both cases we set $\lambda_\text{SO}=30$ meV and let $\varepsilon_c$ change from $0$ to 
$\varepsilon_h$. 

\begin{figure}[!htp]
	\centering
	\subfloat[]{\includegraphics[width=8.3cm]{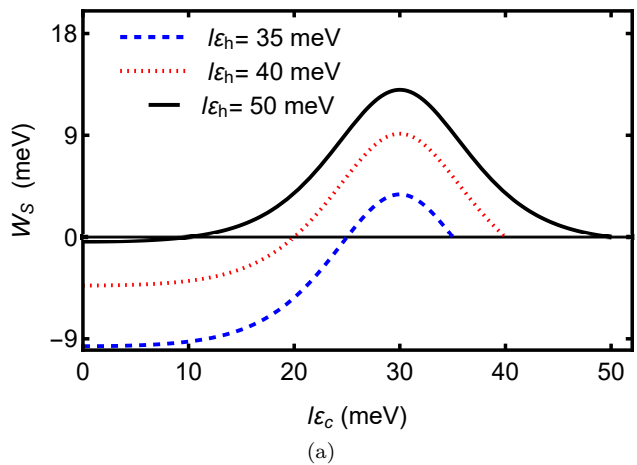}\label{fig:2levelwork}}\\
	\subfloat[]{\includegraphics[width=8.3cm]{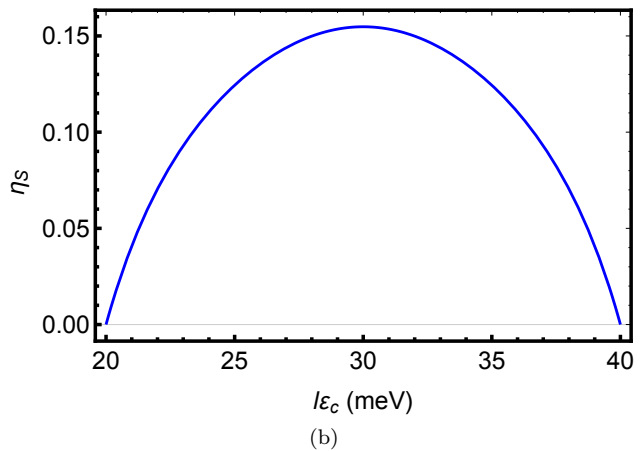} \label{fig:2leveleff}}
	\caption{{(Color Online)(a) Work  $W_{\text{S}}$ of two-level Sn model under Stirling cycle, as a function of the electric potential at the cold isoelectric stage $l\varepsilon_{c}$. Spin-orbit coupling strength is  $\lambda_\text{SO}=30 meV$, hot and cold baths are at temperatures
			$T_h=40$ K and $T_{c}=30$ K and the electric potential at the hot isoelectric stage is $l\varepsilon_h=35$ meV (blue long dashed), $l\varepsilon_h=40$ meV (red short dashed) and $l\varepsilon_h=50$ meV (black solid)  (b) Efficiency $\eta_{\text{S}}$ of this system for  $l\varepsilon_h=40$ meV. }}
	\label{fig:4band}
\end{figure}

In the low temperature case, the temperatures of the hot and cold baths are $T_h=40$ K and $T_c=30$ K, respectively.  Fig.~\ref{fig:4band} displays the net work and the efficiency of the cycle for the low temperature case. 
In Fig.~\ref{fig:2levelwork} it is observed that for different values of $l\varepsilon_{h}$ the maximum value of work occur at $\lambda_\text{SO}= l\varepsilon_c$. These peaks identify the critical point of TPT. 
 Increasing the magnitude of electric field causes a broader region of positive work.
Fig.~\ref{fig:2leveleff} which shows the efficiency of this cycle for $l\varepsilon_h$=40 meV, is a completely symmetric plot with a maximum at TPT point. 

\begin{figure}[!htp]
\centering
\includegraphics[width=8.3cm]{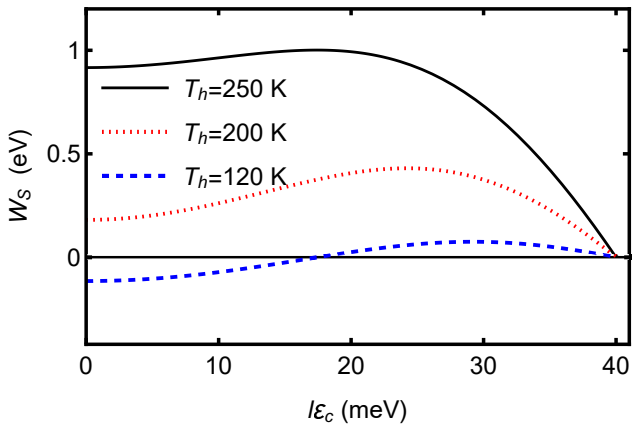}

\caption{{(Color Online) Work $W_{\text{S}}$  of  Stirling cycle as a function of the electric potential at the cold isochore $l\varepsilon_{c}$. The electric potential at the hot isoelectric is $l\varepsilon_h=40$ meV and the spin-orbit coupling strength is $\lambda_{\text{SO}}=30$ meV. Cold bath is at temperatures
 $T_{c}=80$ K and the hot both is at $T_{h}=120$ K (blue long dashed), $T_{c}=200$ K (red short dashed) and $T_{c}=250 $ K (black solid).}}
\label{fig:stirhigh}
\end{figure}

For the high temperature case, the behavior of work output as a function of $l\varepsilon_{c}$, is plotted in Fig.~\ref{fig:stirhigh} for three different values of $T_h$. As the temperature of the hot bath is increased, the maximum of work is shifted away from TPT and the sign of TPT point is smoothed out.  


\section{Implementations of the thermodynamic cycles with Topological Insulators}\label{sec:experiment}
In addition to Sn (or Germanene or Silicene) as the working substance, the basic ingredients needed to construct a room temperature quantum heat engine with a TPT are tunable hot and cold
baths that would be periodically in contact with the monolayer TI and tunable electric field that will vary between two values above and below
the critical point. There are experimental reports on the successful fabrication of such monolayer TIs ~\cite{zqh,swz,bke,cmj}. The effect of an external electric field on them has been studied in both theoretical and experimental works~\cite{murakami2011gap,moore2007topological}. While one may directly use the environment (room) temperature, more controllable environments can also be envisioned for monolayer TIs. For example, we may propose that a graphene bilayer can be used as a scaffold for Sn. Such a set up preserves the topological properties of stanene and moreover, these stacked layers are stable above room temperature~\cite{murakami2011gap,moore2007topological}. Graphene has already been studied for its significant potential for heat flow control and energy harvesting~\cite{murakami2011gap,moore2007topological}; and there are investigations which deal with annealing graphene~\cite{murakami2011gap,moore2007topological}. Accordingly, graphene bilayer scaffold could be used to conduct heat in a controllable manner
and can act as a tunable act as the hot and cold bath to implement the isothermal or isochoric heating and cooling stages of the cycles we consider.  
 An external electric field is required for the operation of the thermodynamic cycles. This can be introduced by
a capacitor. 

Similar to typical experimental quantum heat engines~\cite{Klaers} or proposed 
graphene flake engine~\cite{mdq}, our system neither stores the work nor 
transfers it to a work source. By including the interaction term for a work source or a load to
Sn model, such extraction of work could be studied. In experiments work is determined indirectly, 
from the injected and received heat. In our case, heat can be
supplied through the bottom substrate to the Sn film via electric currents with thermal noise, which 
could be measured to determine the work
associated with the Otto and Stirling cycles.
\section{conclusions}\label{sec:conc}

We have investigated signatures of the topological phase transition (TPT) of a monolayer Stanene (Sn) under an applied electric field in the work and efficiency of 
thermodynamic engine cycles using the Sn as their working substance. We specifically considered an Otto cycle and a Stirling cycle. In the case of Otto cycle, positive work domain in the electric field space, as well as the magnitude of the work and efficiency, exhibit qualitative effects of TPT. Depending on the applied fields, either a local minima separates
a double-peaked work output profile or work output is maximum at the critical point of TPT.
In Stirling cycle at low temperature, the work achieves a maximum at the TPT point: while
in contrast to Otto cycle, at high temperature, the maximum shifts away from the TPT point and the peak is smoothed out.
This result generalizes the idea of the Ref.~\cite{qtcw}, which
is using heat engine cycles to probe ordinary quantum
phase transitions (QPTs) to the probing topological QPTs.
Work output is greatly enhanced as the engine operates at higher temperatures in both cycles. The system can operate either as a quantum heat engine or a refrigerator at both low and high temperature regimes and the type of operation can be tuned by a wide range of applied electric fields.

Our results are applicable to other monolayer topological insulators, such as Germanene and Silicene. 
The crucial step here is that it is established for these materials that bulk gap closing (under electric field variation) is associated with a topological phase transition.
Our method can be an alternative to the typical scheme of probing
TPT by the determination of the work distribution under a sudden quench,
if it is known that bulk gap closing reflects the
	topological transition of the system.

The advantage of large spin-orbit coupling in such materials, relative to recently proposed graphene flake heat engines~\cite{mdq}, is to
enhance the work and efficiency with SO coupling and room or higher temperature operation. 
We hope our results could inspire further studies of topological quantum heat engines. 
 
\section*{Acknowledgements}
We acknowledge support by the Scientific and Technological Research Council of Turkey (T\"{U}B{\.I}TAK), 
Grant No.~ (117F097) and by the EU-COST Action (CA16221).
\appendix
\section{Work Expression for Quantum Otto Cycle }\label{app:A}
The low-energy effective Hamiltonian $H^s_{\eta}$ for stanene has the following matrix form:
$$
\small{
 \begin{pmatrix} \eta  \lambda_{\text{SO}}+l\varepsilon_z& 0 & k_x+i\eta k_y & 0\\ 0&-\eta \lambda_{\text{SO}}+l\varepsilon_z&0&k_x+i\eta k_y\\
 k_x-i\eta k_y&0&-\eta \lambda_{\text{SO}}-l\varepsilon_z &0\\ 0&k_x-i\eta k_y&0&\eta \lambda_{\text{SO}}-l\varepsilon_z \end{pmatrix}}$$
which results in the eigenenergies given by
\begin{eqnarray}
E^{-}(k)&=&\pm\sqrt{k^2+(l\varepsilon_z-\eta\lambda_\text{SO})^2},\label{eetas1} \\ 
E^{+}(k)&=&\pm\sqrt{k^2+(l\varepsilon_z+\eta\lambda_\text{SO})^2}.
\label{eetas2}
\end{eqnarray}
Here $\eta=\pm 1$ denotes the $K$ and $K'$ valleys, respectively. Plugging in $\eta$, both Eq.~(\ref{eetas1}) and Eq.~(\ref{eetas2}) result in the set of four distinct energy levels, which can be expressed in a compact form as in Eq.~(\ref{eq:energyseta}), where each energy level is two-fold degenerate.
The upper two bands above the Fermi level corresponds to two positive energies given in Eq.~(\ref{eq:energyseta}):
\begin{eqnarray}
E_1^i(k)&=&\sqrt{k^2+|l\varepsilon_i-\lambda_\text{SO}|^2},\nonumber \\ 
E_2^i(k)&=&\sqrt{k^2+(l\varepsilon_i+\lambda_\text{SO})^2},
\label{e0}
\end{eqnarray}
with $i=h,c$. We set $\hbar=v_f=1.$
The degeneracy factor is  $g(E_n^i)=2$ for each $E^i_n$, where $n=1,2$. The Fermi distribution function for neutral Sn at points $B$ and $D$ of the quantum Otto cycle are given by
\begin{eqnarray}
f_n(B)&=&f_n(k,\varepsilon_h,T_h)=\frac{1}{e^{\beta_hE^h_n(k)}+1}, \nonumber\\
f_n(D)&=&f_n(k,\varepsilon_c,T_c)=\frac{1}{e^{\beta_cE^c_n(k)}+1}.
\end{eqnarray}

Due to the probability conservation on the adiabatic branches of the Otto cycle, we have $f_n(B)=f_n(k,\varepsilon_h,T_h)=f_n(C)$ and $f_n(D)=f_n(k,\varepsilon_c,T_c)=f_n(A).$

The heat received $Q_{\text{in}}$ and ejected $Q_{\text{out}}$, are then obtained as
\begin{eqnarray}
Q^{+}_{\text{in}}=&&\int \frac{d^2\vec k}{(2\pi)^2}\sum^2_{n=1}2E_n^h(k)[f_n(k,\varepsilon_h,T_h)-f_n(k,\varepsilon_c,T_c)]\nonumber\\
=&&\int_{0}^{\infty}dk\frac{k}{\pi}\times\nonumber\\
&&\ \ \ \ \ \ \  \Big[E_1^h\big(f_1(B)-f_1(D)\big)+E_2^h\big(f_2(B)-f_2(D)\big)\Big]\nonumber\\
\label{eq:qinexpl}
\end{eqnarray}
and
\begin{eqnarray}
Q^{+}_{\text{out}} =&&\int \frac{d^2\vec k}{(2\pi)^2}\sum^2_{n=1}2E_n^c(k)[f_n(k,\varepsilon_c,T_c)-f_n(k,\varepsilon_h,T_h)]\nonumber\\
=&&\int_{0}^{\infty}dk\frac{k}{\pi}\times\nonumber\\
&&\ \ \ \ \ \ \  \Big[E_1^c\big(f_1(D)-f_1(B)\big)+E_2^c\big(f_2(D)-f_2(B)\big)\Big]\nonumber\\
\label{eq:qoutexpl}
\end{eqnarray}
respectively. Superscript $+$ of $Q$ indicates that only the positive energy bands are used in the calculation. By using the integrals given in Eq.~(\ref{eq:qinexpl}) and Eq.~(\ref{eq:qoutexpl}) in Eq.~(\ref{eq:work}), contribution of the positive energy bands to the work output can be calculated; we denote it by $W^{+}_\text{O}$.
	Let us now take into consideration the negative energy bands; which can be written as
	\begin{eqnarray}
	E_1^{i(-)}(k)&=&-E_1^i(k),\nonumber \\ 
	E_2^{i(-)}(k)&=&-E_2^i(k).
	\label{eneg}
	\end{eqnarray}
	The average occupation number of these bands is given by
	\begin{eqnarray}
	f^{(-)}_n(k,\varepsilon_i,T_i)=1-f_n(k,\varepsilon_i,T_i).
	\label{fneg}
	\end{eqnarray}
	Directly using the negative energy bands in the heat transfer calculation would lead to divergencies. We use a simple renormalization of these
	infinities as described below.
	
	The ground state energy of the system is determined by $\sum_{n=1}^{2}E_n^{(-)}$  at $T=0$. It is not bounded from below. Nevertheless we can formally subtract it from the internal energy of the Sn 
	at each point in the engine cycle. Such a simple renormalization will not change the value of the net work output, as the additional terms will cancel each other in the work
	calculation, 
	but allows for finding finite energies at each cycle
	point.
	The internal energy $U(I)$ at a specific point $I$ in the cycle then becomes
	\begin{eqnarray}
	U(I)&=&\sum_{n=1}^{2}f_nE_n+\sum_{n=1}^{2}f_n^{(-)}E_n^{(-)}-\sum_{n=1}^{2}E_n^{(-)},\nonumber \\ 
	&=&2\sum_{n=1}^{2}f_nE_n.
	\label{justified}
	\end{eqnarray}
	For notational simplicity, the integral over $k$, superscripts, and the arguments associated with the point $I$ in $f_n$ and $E(k)$ are suppressed.
	The result in Eq.~(\ref{justified}) is obtained in the light of Eq.~(\ref{eneg}) and Eq.~(\ref{fneg}). Accordingly, work output
	of the cycle can be determined by using the relation $W_\text{O}=2W^{+}_\text{O}$.

\section{Work Expression for Stirling Cycle}\label{app:B}
At each point in the Stirling cycle given in Fig.~\ref{fig:cycle2}, we evaluate the Fermi-Dirac distribution as
\begin{eqnarray}
f_n(A)&=&f_n(k,T_h,\varepsilon_h),\\
f_n(B)&=&f_n(k,T_h,\varepsilon_c),\\
f_n(C)&=&f_n(k,T_c,\varepsilon_c),\\
f_n(D)&=&f_n(k,T_c,\varepsilon_h).
\end{eqnarray}
The incoming and outgoing heat are
\begin{eqnarray}
Q_{\text{in}}&=&Q_{BA}+Q_{AD}=T_h\big(S(B)-S(A)\big)+U(D)-U(A)\nonumber\\ \\
Q_{\text{out}}&=&Q_{CB}+Q_{DC}=U(C)-U(B)+T_c\big(S(D)-S(C)\big)\nonumber\\
\end{eqnarray}
In order to calculate work $W_{\text{S}}=Q_{\text{in}}+Q_{\text{out}}$, we need to compute quantities like $T_jS(I)-U(I)$, using
\begin{eqnarray}
U^{+}(I)&=&2\int\frac{d^2\vec k}{(2\pi)^2}\sum_{n=1}^{2} E^i_{n}(k)f_n(I),
\label{internal}
\end{eqnarray}
with $E^i_{n}(k)$ being the corresponding eigenenergy at $I$ and
\begin{eqnarray}
T_jS^{+}(I)&&=-\frac{2}{\beta_j}\int\frac{d^2\vec k}{(2\pi)^2}\times\nonumber\\
&&\sum_{n=1}^{2} \Big[\big(1-f_n(I)\big)\ln
\big(1-f_n(I)\big)
+f_n(I)\ln
\big(f_n(I)\big)\Big].\nonumber\\
\label{entropy}
\end{eqnarray}
Again the superscripts remind us that we limit ourselves to the positive energy manifold in these calculations.
Eqs.~(\ref{internal}) and (\ref{entropy}) yield
\begin{eqnarray}
T_jS^{+}(I)-U^{+}(I)=\frac{2}{\beta_j}\int\frac{d^2 \vec k}{(2\pi)^2}\sum_{n=1}^{2}\ln(1+e^{-\beta_jE_n^i(k)}).\nonumber\\
\label{keyexpres}
\end{eqnarray}
Using Eq.~(\ref{keyexpres}) in Eq.~(\ref{ws1}), we determine the work associated with the positive energy bands and denote it as $W^{+}_{\text{S}}.$ The work expression for the Stirling cycle given in Eq.~(\ref{wstirling}) is valid for any number of $n$.

 In order to include the effect of the negative energy manifold we first realize that $T_jS^{+}(I)=T_jS^{-}(I)$. Then we can employ the
same method, described in Appendix~\ref{app:A} to renormalize the divergent contribution of negative energies to the internal energy $U(I)$ 
by substracting the zero temperature ground state energy at each point $I$ in the Stirling cycle. This leads to $U(I)=2U^+(I)$. 
Hence we conclude that inclusion of the negative energies simply doubles the value of $T_jS^+(I)-U^+(I)$, and accordingly we have 
$W_{\text{S}}=2W^{+}_{\text{S}}.$ We remark a subtle technical point that the ground state energy term formally depends on the associated electric potential at the corresponding point $I$. Thus, we should designate the ground state energy terms substracted from the internal energies at points $A$ and $D$ of the Stirling cycle by $R(\varepsilon_{h})$, and at points $B$ and $C$ by $R(\varepsilon_{c})$. The heat exchanged at the isoelectric branches including these extra terms can then be expressed as
\begin{eqnarray}
Q_{BA}&=&T_h(S(B)-S(A))-R(\varepsilon_{c})+R(\varepsilon_{h}), \label{qba4} \\
Q_{DC}&=&T_c(S(D)-S(C))-R(\varepsilon_{h})+R(\varepsilon_{c}), \label{qdc4}
\end{eqnarray}
from which we see that the extra terms cancel properly in the calculation of the net work output of the cycle by~Eq.(\ref{ws1}).
   

\begin{thebibliography}{55}%
	\makeatletter
	\providecommand \@ifxundefined [1]{%
		\@ifx{#1\undefined}
	}%
	\providecommand \@ifnum [1]{%
		\ifnum #1\expandafter \@firstoftwo
		\else \expandafter \@secondoftwo
		\fi
	}%
	\providecommand \@ifx [1]{%
		\ifx #1\expandafter \@firstoftwo
		\else \expandafter \@secondoftwo
		\fi
	}%
	\providecommand \natexlab [1]{#1}%
	\providecommand \enquote  [1]{``#1''}%
	\providecommand \bibnamefont  [1]{#1}%
	\providecommand \bibfnamefont [1]{#1}%
	\providecommand \citenamefont [1]{#1}%
	\providecommand \href@noop [0]{\@secondoftwo}%
	\providecommand \href [0]{\begingroup \@sanitize@url \@href}%
	\providecommand \@href[1]{\@@startlink{#1}\@@href}%
	\providecommand \@@href[1]{\endgroup#1\@@endlink}%
	\providecommand \@sanitize@url [0]{\catcode `\\12\catcode `\$12\catcode
		`\&12\catcode `\#12\catcode `\^12\catcode `\_12\catcode `\%12\relax}%
	\providecommand \@@startlink[1]{}%
	\providecommand \@@endlink[0]{}%
	\providecommand \url  [0]{\begingroup\@sanitize@url \@url }%
	\providecommand \@url [1]{\endgroup\@href {#1}{\urlprefix }}%
	\providecommand \urlprefix  [0]{URL }%
	\providecommand \Eprint [0]{\href }%
	\providecommand \doibase [0]{http://dx.doi.org/}%
	\providecommand \selectlanguage [0]{\@gobble}%
	\providecommand \bibinfo  [0]{\@secondoftwo}%
	\providecommand \bibfield  [0]{\@secondoftwo}%
	\providecommand \translation [1]{[#1]}%
	\providecommand \BibitemOpen [0]{}%
	\providecommand \bibitemStop [0]{}%
	\providecommand \bibitemNoStop [0]{.\EOS\space}%
	\providecommand \EOS [0]{\spacefactor3000\relax}%
	\providecommand \BibitemShut  [1]{\csname bibitem#1\endcsname}%
	\let\auto@bib@innerbib\@empty
	\bibitem [{\citenamefont {Cardwell}(1973)}]{fwt}%
	\BibitemOpen
	\bibfield  {author} {\bibinfo {author} {\bibfnamefont {D.~S.~L.}\
			\bibnamefont {Cardwell}},\ }\bibfield  {title} {\enquote {\bibinfo {title}
			{From watt to clausius: The rise of thermodynamics in the early industrial
				age. d. s. l. cardwell},}\ }\href {\doibase 10.1086/620135} {\bibfield
		{journal} {\bibinfo  {journal} {The Library Quarterly}\ }\textbf {\bibinfo
			{volume} {43}},\ \bibinfo {pages} {168--169} (\bibinfo {year}
		{1973})}\BibitemShut {NoStop}%
	\bibitem [{\citenamefont {Scovil}\ and\ \citenamefont
		{{Schulz-DuBois}}(1959)}]{tlm}%
	\BibitemOpen
	\bibfield  {author} {\bibinfo {author} {\bibfnamefont {H.~E.~D.}\
			\bibnamefont {Scovil}}\ and\ \bibinfo {author} {\bibfnamefont {E.~O.}\
			\bibnamefont {{Schulz-DuBois}}},\ }\bibfield  {title} {\enquote {\bibinfo
			{title} {Three-{{Level Masers}} as {{Heat Engines}}},}\ }\href {\doibase
		10.1103/PhysRevLett.2.262} {\bibfield  {journal} {\bibinfo  {journal} {Phys.
				Rev. Lett.}\ }\textbf {\bibinfo {volume} {2}},\ \bibinfo {pages} {262--263}
		(\bibinfo {year} {1959})}\BibitemShut {NoStop}%
	\bibitem [{\citenamefont {Scully}\ \emph {et~al.}(2003)\citenamefont {Scully},
		\citenamefont {Zubairy}, \citenamefont {Agarwal},\ and\ \citenamefont
		{Walther}}]{ewf}%
	\BibitemOpen
	\bibfield  {author} {\bibinfo {author} {\bibfnamefont {M.~O.}\ \bibnamefont
			{Scully}}, \bibinfo {author} {\bibfnamefont {M.~S.}\ \bibnamefont {Zubairy}},
		\bibinfo {author} {\bibfnamefont {G.~S.}\ \bibnamefont {Agarwal}}, \ and\
		\bibinfo {author} {\bibfnamefont {H.}~\bibnamefont {Walther}},\ }\bibfield
	{title} {\enquote {\bibinfo {title} {Extracting {{Work}} from a {{Single Heat
						Bath}} via {{Vanishing Quantum Coherence}}},}\ }\href {\doibase
		10.1126/science.1078955} {\bibfield  {journal} {\bibinfo  {journal}
			{Science}\ }\textbf {\bibinfo {volume} {299}},\ \bibinfo {pages} {862--864}
		(\bibinfo {year} {2003})}\BibitemShut {NoStop}%
	\bibitem [{\citenamefont {Ryan}\ \emph {et~al.}(2008)\citenamefont {Ryan},
		\citenamefont {Moussa}, \citenamefont {Baugh},\ and\ \citenamefont
		{Laflamme}}]{dom}%
	\BibitemOpen
	\bibfield  {author} {\bibinfo {author} {\bibfnamefont {C.~A.}\ \bibnamefont
			{Ryan}}, \bibinfo {author} {\bibfnamefont {O.}~\bibnamefont {Moussa}},
		\bibinfo {author} {\bibfnamefont {J.}~\bibnamefont {Baugh}}, \ and\ \bibinfo
		{author} {\bibfnamefont {R.}~\bibnamefont {Laflamme}},\ }\bibfield  {title}
	{\enquote {\bibinfo {title} {Spin {{Based Heat Engine}}: {{Demonstration}} of
				{{Multiple Rounds}} of {{Algorithmic Cooling}}},}\ }\href {\doibase
		10.1103/PhysRevLett.100.140501} {\bibfield  {journal} {\bibinfo  {journal}
			{Phys. Rev. Lett.}\ }\textbf {\bibinfo {volume} {100}},\ \bibinfo {pages}
		{140501} (\bibinfo {year} {2008})}\BibitemShut {NoStop}%
	\bibitem [{\citenamefont {Ro\ss{}nagel}\ \emph {et~al.}(2016)\citenamefont
		{Ro\ss{}nagel}, \citenamefont {Dawkins}, \citenamefont {Tolazzi},
		\citenamefont {Abah}, \citenamefont {Lutz}, \citenamefont {{Schmidt-Kaler}},\
		and\ \citenamefont {Singer}}]{asa}%
	\BibitemOpen
	\bibfield  {author} {\bibinfo {author} {\bibfnamefont {Johannes}\
			\bibnamefont {Ro\ss{}nagel}}, \bibinfo {author} {\bibfnamefont {Samuel~T.}\
			\bibnamefont {Dawkins}}, \bibinfo {author} {\bibfnamefont {Karl~N.}\
			\bibnamefont {Tolazzi}}, \bibinfo {author} {\bibfnamefont {Obinna}\
			\bibnamefont {Abah}}, \bibinfo {author} {\bibfnamefont {Eric}\ \bibnamefont
			{Lutz}}, \bibinfo {author} {\bibfnamefont {Ferdinand}\ \bibnamefont
			{{Schmidt-Kaler}}}, \ and\ \bibinfo {author} {\bibfnamefont {Kilian}\
			\bibnamefont {Singer}},\ }\bibfield  {title} {\enquote {\bibinfo {title} {A
				single-atom heat engine},}\ }\href {\doibase 10.1126/science.aad6320}
	{\bibfield  {journal} {\bibinfo  {journal} {Science}\ }\textbf {\bibinfo
			{volume} {352}},\ \bibinfo {pages} {325--329} (\bibinfo {year}
		{2016})}\BibitemShut {NoStop}%
	\bibitem [{\citenamefont {Hardal}\ and\ \citenamefont {M{\"u}stecapl\i{}o{\u
				g}lu}(2015)}]{sqh}%
	\BibitemOpen
	\bibfield  {author} {\bibinfo {author} {\bibfnamefont {Ali {\"U}.~C.}\
			\bibnamefont {Hardal}}\ and\ \bibinfo {author} {\bibfnamefont
			{{\"O}zg{\"u}r~E.}\ \bibnamefont {M{\"u}stecapl\i{}o{\u g}lu}},\ }\bibfield
	{title} {\enquote {\bibinfo {title} {Superradiant {{Quantum Heat Engine}}},}\
	}\href {\doibase 10.1038/srep12953} {\bibfield  {journal} {\bibinfo
			{journal} {Scientific Reports}\ }\textbf {\bibinfo {volume} {5}},\ \bibinfo
		{pages} {12953} (\bibinfo {year} {2015})}\BibitemShut {NoStop}%
	\bibitem [{\citenamefont {Zhang}\ \emph {et~al.}(2014)\citenamefont {Zhang},
		\citenamefont {Bariani},\ and\ \citenamefont {Meystre}}]{qoh}%
	\BibitemOpen
	\bibfield  {author} {\bibinfo {author} {\bibfnamefont {Keye}\ \bibnamefont
			{Zhang}}, \bibinfo {author} {\bibfnamefont {Francesco}\ \bibnamefont
			{Bariani}}, \ and\ \bibinfo {author} {\bibfnamefont {Pierre}\ \bibnamefont
			{Meystre}},\ }\bibfield  {title} {\enquote {\bibinfo {title} {Quantum
				{{Optomechanical Heat Engine}}},}\ }\href {\doibase
		10.1103/PhysRevLett.112.150602} {\bibfield  {journal} {\bibinfo  {journal}
			{Phys. Rev. Lett.}\ }\textbf {\bibinfo {volume} {112}},\ \bibinfo {pages}
		{150602} (\bibinfo {year} {2014})}\BibitemShut {NoStop}%
	\bibitem [{\citenamefont {Hardal}\ \emph {et~al.}(2017)\citenamefont {Hardal},
		\citenamefont {Aslan}, \citenamefont {Wilson},\ and\ \citenamefont
		{M{\"u}stecapl\i{}o{\u g}lu}}]{qhew}%
	\BibitemOpen
	\bibfield  {author} {\bibinfo {author} {\bibfnamefont {Ali {\"U}.~C.}\
			\bibnamefont {Hardal}}, \bibinfo {author} {\bibfnamefont {Nur}\ \bibnamefont
			{Aslan}}, \bibinfo {author} {\bibfnamefont {C.~M.}\ \bibnamefont {Wilson}}, \
		and\ \bibinfo {author} {\bibfnamefont {{\"O}zg{\"u}r~E.}\ \bibnamefont
			{M{\"u}stecapl\i{}o{\u g}lu}},\ }\bibfield  {title} {\enquote {\bibinfo
			{title} {Quantum heat engine with coupled superconducting resonators},}\
	}\href {\doibase 10.1103/PhysRevE.96.062120} {\bibfield  {journal} {\bibinfo
			{journal} {Phys. Rev. E}\ }\textbf {\bibinfo {volume} {96}},\ \bibinfo
		{pages} {062120} (\bibinfo {year} {2017})}\BibitemShut {NoStop}%
	\bibitem [{\citenamefont {Pe{\~n}a}\ \emph {et~al.}(2016)\citenamefont
		{Pe{\~n}a}, \citenamefont {Ferr{\'e}}, \citenamefont {Orellana},
		\citenamefont {Rojas},\ and\ \citenamefont {Vargas}}]{ooa}%
	\BibitemOpen
	\bibfield  {author} {\bibinfo {author} {\bibfnamefont {Francisco~J.}\
			\bibnamefont {Pe{\~n}a}}, \bibinfo {author} {\bibfnamefont {Michel}\
			\bibnamefont {Ferr{\'e}}}, \bibinfo {author} {\bibfnamefont {P.~A.}\
			\bibnamefont {Orellana}}, \bibinfo {author} {\bibfnamefont {Ren{\'e}~G.}\
			\bibnamefont {Rojas}}, \ and\ \bibinfo {author} {\bibfnamefont
			{P.}~\bibnamefont {Vargas}},\ }\bibfield  {title} {\enquote {\bibinfo {title}
			{Optimization of a relativistic quantum mechanical engine},}\ }\href
	{\doibase 10.1103/PhysRevE.94.022109} {\bibfield  {journal} {\bibinfo
			{journal} {Phys. Rev. E}\ }\textbf {\bibinfo {volume} {94}},\ \bibinfo
		{pages} {022109} (\bibinfo {year} {2016})}\BibitemShut {NoStop}%
	\bibitem [{\citenamefont {Mu{\~n}oz}\ and\ \citenamefont
		{Pe{\~n}a}(2012)}]{qhei}%
	\BibitemOpen
	\bibfield  {author} {\bibinfo {author} {\bibfnamefont {Enrique}\ \bibnamefont
			{Mu{\~n}oz}}\ and\ \bibinfo {author} {\bibfnamefont {Francisco~J.}\
			\bibnamefont {Pe{\~n}a}},\ }\bibfield  {title} {\enquote {\bibinfo {title}
			{Quantum heat engine in the relativistic limit: {{The}} case of a {{Dirac}}
				particle},}\ }\href {\doibase 10.1103/PhysRevE.86.061108} {\bibfield
		{journal} {\bibinfo  {journal} {Phys. Rev. E}\ }\textbf {\bibinfo {volume}
			{86}},\ \bibinfo {pages} {061108} (\bibinfo {year} {2012})}\BibitemShut
	{NoStop}%
	\bibitem [{\citenamefont {Pe{\~n}a}\ and\ \citenamefont
		{Mu{\~n}oz}(2015)}]{mdq}%
	\BibitemOpen
	\bibfield  {author} {\bibinfo {author} {\bibfnamefont {Francisco~J.}\
			\bibnamefont {Pe{\~n}a}}\ and\ \bibinfo {author} {\bibfnamefont {Enrique}\
			\bibnamefont {Mu{\~n}oz}},\ }\bibfield  {title} {\enquote {\bibinfo {title}
			{Magnetostrain-driven quantum engine on a graphene flake},}\ }\href {\doibase
		10.1103/PhysRevE.91.052152} {\bibfield  {journal} {\bibinfo  {journal} {Phys.
				Rev. E}\ }\textbf {\bibinfo {volume} {91}},\ \bibinfo {pages} {052152}
		(\bibinfo {year} {2015})}\BibitemShut {NoStop}%
	\bibitem [{\citenamefont {Mu{\~n}oz}\ \emph {et~al.}(2016)\citenamefont
		{Mu{\~n}oz}, \citenamefont {Pe{\~n}a},\ and\ \citenamefont
		{Gonz{\'a}lez}}]{mdqh}%
	\BibitemOpen
	\bibfield  {author} {\bibinfo {author} {\bibfnamefont {Enrique}\ \bibnamefont
			{Mu{\~n}oz}}, \bibinfo {author} {\bibfnamefont {Francisco~J.}\ \bibnamefont
			{Pe{\~n}a}}, \ and\ \bibinfo {author} {\bibfnamefont {Alejandro}\
			\bibnamefont {Gonz{\'a}lez}},\ }\bibfield  {title} {\enquote {\bibinfo
			{title} {Magnetically-{{Driven Quantum Heat Engines}}: {{The Quasi}}-{{Static
						Limit}} of {{Their Efficiency}}},}\ }\href {\doibase 10.3390/e18050173}
	{\bibfield  {journal} {\bibinfo  {journal} {Entropy}\ }\textbf {\bibinfo
			{volume} {18}},\ \bibinfo {pages} {173} (\bibinfo {year} {2016})}\BibitemShut
	{NoStop}%
	\bibitem [{\citenamefont {Mani}\ and\ \citenamefont {Benjamin}(2017)}]{sgb}%
	\BibitemOpen
	\bibfield  {author} {\bibinfo {author} {\bibfnamefont {Arjun}\ \bibnamefont
			{Mani}}\ and\ \bibinfo {author} {\bibfnamefont {Colin}\ \bibnamefont
			{Benjamin}},\ }\bibfield  {title} {\enquote {\bibinfo {title}
			{Strained-graphene-based highly efficient quantum heat engine operating at
				maximum power},}\ }\href {\doibase 10.1103/PhysRevE.96.032118} {\bibfield
		{journal} {\bibinfo  {journal} {Phys Rev E}\ }\textbf {\bibinfo {volume}
			{96}},\ \bibinfo {pages} {032118} (\bibinfo {year} {2017})}\BibitemShut
	{NoStop}%
	\bibitem [{\citenamefont {Johnson}(2014)}]{hhe}%
	\BibitemOpen
	\bibfield  {author} {\bibinfo {author} {\bibfnamefont {Clifford~V}\
			\bibnamefont {Johnson}},\ }\bibfield  {title} {\enquote {\bibinfo {title}
			{Holographic heat engines},}\ }\href {\doibase
		10.1088/0264-9381/31/20/205002} {\bibfield  {journal} {\bibinfo  {journal}
			{Class. Quantum Grav.}\ }\textbf {\bibinfo {volume} {31}},\ \bibinfo {pages}
		{205002} (\bibinfo {year} {2014})}\BibitemShut {NoStop}%
	\bibitem [{\citenamefont {Brantut}\ \emph {et~al.}(2013)\citenamefont
		{Brantut}, \citenamefont {Grenier}, \citenamefont {Meineke}, \citenamefont
		{Stadler}, \citenamefont {Krinner}, \citenamefont {Kollath}, \citenamefont
		{Esslinger},\ and\ \citenamefont {Georges}}]{ath}%
	\BibitemOpen
	\bibfield  {author} {\bibinfo {author} {\bibfnamefont {Jean-Philippe}\
			\bibnamefont {Brantut}}, \bibinfo {author} {\bibfnamefont {Charles}\
			\bibnamefont {Grenier}}, \bibinfo {author} {\bibfnamefont {Jakob}\
			\bibnamefont {Meineke}}, \bibinfo {author} {\bibfnamefont {David}\
			\bibnamefont {Stadler}}, \bibinfo {author} {\bibfnamefont {Sebastian}\
			\bibnamefont {Krinner}}, \bibinfo {author} {\bibfnamefont {Corinna}\
			\bibnamefont {Kollath}}, \bibinfo {author} {\bibfnamefont {Tilman}\
			\bibnamefont {Esslinger}}, \ and\ \bibinfo {author} {\bibfnamefont {Antoine}\
			\bibnamefont {Georges}},\ }\bibfield  {title} {\enquote {\bibinfo {title} {A
				{{Thermoelectric Heat Engine}} with {{Ultracold Atoms}}},}\ }\href {\doibase
		10.1126/science.1242308} {\bibfield  {journal} {\bibinfo  {journal}
			{Science}\ }\textbf {\bibinfo {volume} {342}},\ \bibinfo {pages} {713--715}
		(\bibinfo {year} {2013})}\BibitemShut {NoStop}%
	\bibitem [{\citenamefont {Quan}\ \emph {et~al.}(2007)\citenamefont {Quan},
		\citenamefont {Liu}, \citenamefont {Sun},\ and\ \citenamefont {Nori}}]{qtca}%
	\BibitemOpen
	\bibfield  {author} {\bibinfo {author} {\bibfnamefont {H.~T.}\ \bibnamefont
			{Quan}}, \bibinfo {author} {\bibfnamefont {Yu-xi}\ \bibnamefont {Liu}},
		\bibinfo {author} {\bibfnamefont {C.~P.}\ \bibnamefont {Sun}}, \ and\
		\bibinfo {author} {\bibfnamefont {Franco}\ \bibnamefont {Nori}},\ }\bibfield
	{title} {\enquote {\bibinfo {title} {Quantum thermodynamic cycles and quantum
				heat engines},}\ }\href {\doibase 10.1103/PhysRevE.76.031105} {\bibfield
		{journal} {\bibinfo  {journal} {Phys. Rev. E}\ }\textbf {\bibinfo {volume}
			{76}},\ \bibinfo {pages} {031105} (\bibinfo {year} {2007})}\BibitemShut
	{NoStop}%
	\bibitem [{\citenamefont {Benenti}\ \emph {et~al.}(2017)\citenamefont
		{Benenti}, \citenamefont {Casati}, \citenamefont {Saito},\ and\ \citenamefont
		{Whitney}}]{fao}%
	\BibitemOpen
	\bibfield  {author} {\bibinfo {author} {\bibfnamefont {Giuliano}\
			\bibnamefont {Benenti}}, \bibinfo {author} {\bibfnamefont {Giulio}\
			\bibnamefont {Casati}}, \bibinfo {author} {\bibfnamefont {Keiji}\
			\bibnamefont {Saito}}, \ and\ \bibinfo {author} {\bibfnamefont {Robert~S.}\
			\bibnamefont {Whitney}},\ }\bibfield  {title} {\enquote {\bibinfo {title}
			{Fundamental aspects of steady-state conversion of heat to work at the
				nanoscale},}\ }\href {\doibase 10.1016/j.physrep.2017.05.008} {\bibfield
		{journal} {\bibinfo  {journal} {Phys. Rep.}\ }\textbf {\bibinfo {volume}
			{694}},\ \bibinfo {pages} {1--124} (\bibinfo {year} {2017})}\BibitemShut
	{NoStop}%
	\bibitem [{\citenamefont {Alicki}\ and\ \citenamefont {Kosloff}(2018)}]{itq}%
	\BibitemOpen
	\bibfield  {author} {\bibinfo {author} {\bibfnamefont {Robert}\ \bibnamefont
			{Alicki}}\ and\ \bibinfo {author} {\bibfnamefont {Ronnie}\ \bibnamefont
			{Kosloff}},\ }\bibfield  {title} {\enquote {\bibinfo {title} {Introduction to
				{{Quantum Thermodynamics}}: {{History}} and {{Prospects}}},}\ }\href@noop {}
	{\bibfield  {journal} {\bibinfo  {journal} {arXiv:1801.08314 [quant-ph]}\ }
		(\bibinfo {year} {2018})},\ \Eprint {http://arxiv.org/abs/1801.08314}
	{arXiv:1801.08314 [quant-ph]} \BibitemShut {NoStop}%
	\bibitem [{\citenamefont {Vinjanampathy}\ and\ \citenamefont
		{Anders}(2016)}]{qt}%
	\BibitemOpen
	\bibfield  {author} {\bibinfo {author} {\bibfnamefont {Sai}\ \bibnamefont
			{Vinjanampathy}}\ and\ \bibinfo {author} {\bibfnamefont {Janet}\ \bibnamefont
			{Anders}},\ }\bibfield  {title} {\enquote {\bibinfo {title} {Quantum
				thermodynamics},}\ }\href {\doibase 10.1080/00107514.2016.1201896} {\bibfield
		{journal} {\bibinfo  {journal} {Contemp. Phys.}\ }\textbf {\bibinfo {volume}
			{57}},\ \bibinfo {pages} {545--579} (\bibinfo {year} {2016})}\BibitemShut
	{NoStop}%
	\bibitem [{\citenamefont {Goold}\ \emph {et~al.}(2016)\citenamefont {Goold},
		\citenamefont {Huber}, \citenamefont {Riera}, \citenamefont {del Rio},\ and\
		\citenamefont {Skrzypczyk}}]{tro}%
	\BibitemOpen
	\bibfield  {author} {\bibinfo {author} {\bibfnamefont {John}\ \bibnamefont
			{Goold}}, \bibinfo {author} {\bibfnamefont {Marcus}\ \bibnamefont {Huber}},
		\bibinfo {author} {\bibfnamefont {Arnau}\ \bibnamefont {Riera}}, \bibinfo
		{author} {\bibfnamefont {L{\'\i}dia}\ \bibnamefont {del Rio}}, \ and\
		\bibinfo {author} {\bibfnamefont {Paul}\ \bibnamefont {Skrzypczyk}},\
	}\bibfield  {title} {\enquote {\bibinfo {title} {The role of quantum
				information in thermodynamics\textemdash{}a topical review},}\ }\href
	{\doibase 10.1088/1751-8113/49/14/143001} {\bibfield  {journal} {\bibinfo
			{journal} {J. Phys A: Math. Theor}\ }\textbf {\bibinfo {volume} {49}},\
		\bibinfo {pages} {143001} (\bibinfo {year} {2016})}\BibitemShut {NoStop}%
	\bibitem [{\citenamefont {Klatzow}\ \emph {et~al.}(2017)\citenamefont
		{Klatzow}, \citenamefont {Becker}, \citenamefont {Ledingham}, \citenamefont
		{Weinzetl}, \citenamefont {Kaczmarek}, \citenamefont {Saunders},
		\citenamefont {Nunn}, \citenamefont {Walmsley}, \citenamefont {Uzdin},\ and\
		\citenamefont {Poem}}]{edo}%
	\BibitemOpen
	\bibfield  {author} {\bibinfo {author} {\bibfnamefont {James}\ \bibnamefont
			{Klatzow}}, \bibinfo {author} {\bibfnamefont {Jonas~N.}\ \bibnamefont
			{Becker}}, \bibinfo {author} {\bibfnamefont {Patrick~M.}\ \bibnamefont
			{Ledingham}}, \bibinfo {author} {\bibfnamefont {Christian}\ \bibnamefont
			{Weinzetl}}, \bibinfo {author} {\bibfnamefont {Krzysztof~T.}\ \bibnamefont
			{Kaczmarek}}, \bibinfo {author} {\bibfnamefont {Dylan~J.}\ \bibnamefont
			{Saunders}}, \bibinfo {author} {\bibfnamefont {Joshua}\ \bibnamefont {Nunn}},
		\bibinfo {author} {\bibfnamefont {Ian~A.}\ \bibnamefont {Walmsley}}, \bibinfo
		{author} {\bibfnamefont {Raam}\ \bibnamefont {Uzdin}}, \ and\ \bibinfo
		{author} {\bibfnamefont {Eilon}\ \bibnamefont {Poem}},\ }\bibfield  {title}
	{\enquote {\bibinfo {title} {Experimental demonstration of quantum effects in
				the operation of microscopic heat engines},}\ }\href@noop {} {\bibfield
		{journal} {\bibinfo  {journal} {arXiv:1710.08716}\ ,\ \bibinfo {pages}
			{1710.08716}} (\bibinfo {year} {2017})},\ \Eprint
	{http://arxiv.org/abs/1710.08716} {arXiv:1710.08716} \BibitemShut {NoStop}%
	\bibitem [{\citenamefont {Zou}\ \emph {et~al.}(2017)\citenamefont {Zou},
		\citenamefont {Jiang}, \citenamefont {Mei}, \citenamefont {Guo},\ and\
		\citenamefont {Du}}]{qheu}%
	\BibitemOpen
	\bibfield  {author} {\bibinfo {author} {\bibfnamefont {Yueyang}\ \bibnamefont
			{Zou}}, \bibinfo {author} {\bibfnamefont {Yue}\ \bibnamefont {Jiang}},
		\bibinfo {author} {\bibfnamefont {Yefeng}\ \bibnamefont {Mei}}, \bibinfo
		{author} {\bibfnamefont {Xianxin}\ \bibnamefont {Guo}}, \ and\ \bibinfo
		{author} {\bibfnamefont {Shengwang}\ \bibnamefont {Du}},\ }\bibfield  {title}
	{\enquote {\bibinfo {title} {Quantum {{Heat Engine Using Electromagnetically
						Induced Transparency}}},}\ }\href {\doibase 10.1103/PhysRevLett.119.050602}
	{\bibfield  {journal} {\bibinfo  {journal} {Phys. Rev. Lett.}\ }\textbf
		{\bibinfo {volume} {119}},\ \bibinfo {pages} {050602} (\bibinfo {year}
		{2017})}\BibitemShut {NoStop}%
	\bibitem [{\citenamefont {Campisi}\ and\ \citenamefont {Fazio}(2016)}]{tpo}%
	\BibitemOpen
	\bibfield  {author} {\bibinfo {author} {\bibfnamefont {Michele}\ \bibnamefont
			{Campisi}}\ and\ \bibinfo {author} {\bibfnamefont {Rosario}\ \bibnamefont
			{Fazio}},\ }\bibfield  {title} {\enquote {\bibinfo {title} {The power of a
				critical heat engine},}\ }\href {\doibase 10.1038/ncomms11895} {\bibfield
		{journal} {\bibinfo  {journal} {Nat. Commun.}\ }\textbf {\bibinfo {volume}
			{7}},\ \bibinfo {pages} {11895} (\bibinfo {year} {2016})}\BibitemShut
	{NoStop}%
	\bibitem [{\citenamefont {Ma}\ \emph {et~al.}(2017)\citenamefont {Ma},
		\citenamefont {Su},\ and\ \citenamefont {Sun}}]{qtcw}%
	\BibitemOpen
	\bibfield  {author} {\bibinfo {author} {\bibfnamefont {Yu-Han}\ \bibnamefont
			{Ma}}, \bibinfo {author} {\bibfnamefont {Shan-He}\ \bibnamefont {Su}}, \ and\
		\bibinfo {author} {\bibfnamefont {Chang-Pu}\ \bibnamefont {Sun}},\ }\bibfield
	{title} {\enquote {\bibinfo {title} {Quantum thermodynamic cycle with
				quantum phase transition},}\ }\href {\doibase 10.1103/PhysRevE.96.022143}
	{\bibfield  {journal} {\bibinfo  {journal} {Phys. Rev. E}\ }\textbf {\bibinfo
			{volume} {96}},\ \bibinfo {pages} {022143} (\bibinfo {year}
		{2017})}\BibitemShut {NoStop}%
	\bibitem [{\citenamefont {Mascarenhas}\ \emph {et~al.}(2014)\citenamefont
		{Mascarenhas}, \citenamefont {Bragan{\c c}a}, \citenamefont {Dorner},
		\citenamefont {Fran{\c c}a~Santos}, \citenamefont {Vedral}, \citenamefont
		{Modi},\ and\ \citenamefont {Goold}}]{waq}%
	\BibitemOpen
	\bibfield  {author} {\bibinfo {author} {\bibfnamefont {E.}~\bibnamefont
			{Mascarenhas}}, \bibinfo {author} {\bibfnamefont {H.}~\bibnamefont {Bragan{\c
					c}a}}, \bibinfo {author} {\bibfnamefont {R.}~\bibnamefont {Dorner}}, \bibinfo
		{author} {\bibfnamefont {M.}~\bibnamefont {Fran{\c c}a~Santos}}, \bibinfo
		{author} {\bibfnamefont {V.}~\bibnamefont {Vedral}}, \bibinfo {author}
		{\bibfnamefont {K.}~\bibnamefont {Modi}}, \ and\ \bibinfo {author}
		{\bibfnamefont {J.}~\bibnamefont {Goold}},\ }\bibfield  {title} {\enquote
		{\bibinfo {title} {Work and quantum phase transitions: {{Quantum}}
				latency},}\ }\href {\doibase 10.1103/PhysRevE.89.062103} {\bibfield
		{journal} {\bibinfo  {journal} {Phys. Rev. E}\ }\textbf {\bibinfo {volume}
			{89}},\ \bibinfo {pages} {062103} (\bibinfo {year} {2014})}\BibitemShut
	{NoStop}%
	\bibitem [{\citenamefont {Silva}(2008)}]{sot}%
	\BibitemOpen
	\bibfield  {author} {\bibinfo {author} {\bibfnamefont {Alessandro}\
			\bibnamefont {Silva}},\ }\bibfield  {title} {\enquote {\bibinfo {title}
			{Statistics of the {{Work Done}} on a {{Quantum Critical System}} by
				{{Quenching}} a {{Control Parameter}}},}\ }\href {\doibase
		10.1103/PhysRevLett.101.120603} {\bibfield  {journal} {\bibinfo  {journal}
			{Phys. Rev. Lett.}\ }\textbf {\bibinfo {volume} {101}},\ \bibinfo {pages}
		{120603} (\bibinfo {year} {2008})}\BibitemShut {NoStop}%
	\bibitem [{\citenamefont {Murakami}(2011)}]{murakami2011gap}%
	\BibitemOpen
	\bibfield  {author} {\bibinfo {author} {\bibfnamefont {Shuichi}\ \bibnamefont
			{Murakami}},\ }\bibfield  {title} {\enquote {\bibinfo {title} {Gap closing
				and universal phase diagrams in topological insulators},}\ }\href {\doibase
		10.1016/j.physe.2010.07.043} {\bibfield  {journal} {\bibinfo  {journal}
			{Physica E}\ }\textbf {\bibinfo {volume} {43}},\ \bibinfo {pages} {748--754}
		(\bibinfo {year} {2011})}\BibitemShut {NoStop}%
	\bibitem [{\citenamefont {Ezawa}(2012)}]{ezawa2012topological}%
	\BibitemOpen
	\bibfield  {author} {\bibinfo {author} {\bibfnamefont {Motohiko}\
			\bibnamefont {Ezawa}},\ }\bibfield  {title} {\enquote {\bibinfo {title} {A
				topological insulator and helical zero mode in silicene under an
				inhomogeneous electric field},}\ }\href {\doibase
		10.1088/1367-2630/14/3/033003} {\bibfield  {journal} {\bibinfo  {journal}
			{New J. Phys.}\ }\textbf {\bibinfo {volume} {14}},\ \bibinfo {pages} {033003}
		(\bibinfo {year} {2012})}\BibitemShut {NoStop}%
	\bibitem [{\citenamefont {Xu}\ \emph {et~al.}(2015)\citenamefont {Xu},
		\citenamefont {Tang},\ and\ \citenamefont {Zhang}}]{lgq}%
	\BibitemOpen
	\bibfield  {author} {\bibinfo {author} {\bibfnamefont {Yong}\ \bibnamefont
			{Xu}}, \bibinfo {author} {\bibfnamefont {Peizhe}\ \bibnamefont {Tang}}, \
		and\ \bibinfo {author} {\bibfnamefont {Shou-Cheng}\ \bibnamefont {Zhang}},\
	}\bibfield  {title} {\enquote {\bibinfo {title} {Large-gap quantum spin
				{Hall} states in decorated stanene grown on a substrate},}\ }\href {\doibase
		10.1103/PhysRevB.92.081112} {\bibfield  {journal} {\bibinfo  {journal} {Phys.
				Rev. B}\ }\textbf {\bibinfo {volume} {92}},\ \bibinfo {pages} {081112}
		(\bibinfo {year} {2015})}\BibitemShut {NoStop}%
	\bibitem [{\citenamefont {Xu}\ \emph {et~al.}(2013)\citenamefont {Xu},
		\citenamefont {Yan}, \citenamefont {Zhang}, \citenamefont {Wang},
		\citenamefont {Xu}, \citenamefont {Tang}, \citenamefont {Duan},\ and\
		\citenamefont {Zhang}}]{lgqs}%
	\BibitemOpen
	\bibfield  {author} {\bibinfo {author} {\bibfnamefont {Yong}\ \bibnamefont
			{Xu}}, \bibinfo {author} {\bibfnamefont {Binghai}\ \bibnamefont {Yan}},
		\bibinfo {author} {\bibfnamefont {Hai-Jun}\ \bibnamefont {Zhang}}, \bibinfo
		{author} {\bibfnamefont {Jing}\ \bibnamefont {Wang}}, \bibinfo {author}
		{\bibfnamefont {Gang}\ \bibnamefont {Xu}}, \bibinfo {author} {\bibfnamefont
			{Peizhe}\ \bibnamefont {Tang}}, \bibinfo {author} {\bibfnamefont {Wenhui}\
			\bibnamefont {Duan}}, \ and\ \bibinfo {author} {\bibfnamefont {Shou-Cheng}\
			\bibnamefont {Zhang}},\ }\bibfield  {title} {\enquote {\bibinfo {title}
			{Large-{{Gap Quantum Spin Hall Insulators}} in {{Tin Films}}},}\ }\href
	{\doibase 10.1103/PhysRevLett.111.136804} {\bibfield  {journal} {\bibinfo
			{journal} {Phys. Rev. Lett.}\ }\textbf {\bibinfo {volume} {111}},\ \bibinfo
		{pages} {136804} (\bibinfo {year} {2013})}\BibitemShut {NoStop}%
	\bibitem [{\citenamefont {Cahangirov}\ \emph {et~al.}(2009)\citenamefont
		{Cahangirov}, \citenamefont {Topsakal}, \citenamefont {Akt{\"u}rk},
		\citenamefont {{\c S}ahin},\ and\ \citenamefont {Ciraci}}]{tao}%
	\BibitemOpen
	\bibfield  {author} {\bibinfo {author} {\bibfnamefont {S.}~\bibnamefont
			{Cahangirov}}, \bibinfo {author} {\bibfnamefont {M.}~\bibnamefont
			{Topsakal}}, \bibinfo {author} {\bibfnamefont {E.}~\bibnamefont
			{Akt{\"u}rk}}, \bibinfo {author} {\bibfnamefont {H.}~\bibnamefont {{\c
					S}ahin}}, \ and\ \bibinfo {author} {\bibfnamefont {S.}~\bibnamefont
			{Ciraci}},\ }\bibfield  {title} {\enquote {\bibinfo {title} {Two- and
				{{One}}-{{Dimensional Honeycomb Structures}} of {{Silicon}} and
				{{Germanium}}},}\ }\href {\doibase 10.1103/PhysRevLett.102.236804} {\bibfield
		{journal} {\bibinfo  {journal} {Phys. Rev. Lett.}\ }\textbf {\bibinfo
			{volume} {102}},\ \bibinfo {pages} {236804} (\bibinfo {year}
		{2009})}\BibitemShut {NoStop}%
	\bibitem [{\citenamefont {Liu}\ \emph {et~al.}(2011)\citenamefont {Liu},
		\citenamefont {Jiang},\ and\ \citenamefont {Yao}}]{lee}%
	\BibitemOpen
	\bibfield  {author} {\bibinfo {author} {\bibfnamefont {Cheng-Cheng}\
			\bibnamefont {Liu}}, \bibinfo {author} {\bibfnamefont {Hua}\ \bibnamefont
			{Jiang}}, \ and\ \bibinfo {author} {\bibfnamefont {Yugui}\ \bibnamefont
			{Yao}},\ }\bibfield  {title} {\enquote {\bibinfo {title} {Low-energy
				effective {{Hamiltonian}} involving spin-orbit coupling in silicene and
				two-dimensional germanium and tin},}\ }\href {\doibase
		10.1103/PhysRevB.84.195430} {\bibfield  {journal} {\bibinfo  {journal} {Phys.
				Rev. B}\ }\textbf {\bibinfo {volume} {84}},\ \bibinfo {pages} {195430}
		(\bibinfo {year} {2011})}\BibitemShut {NoStop}%
	\bibitem [{\citenamefont {Fadaie}\ \emph {et~al.}(2017)\citenamefont {Fadaie},
		\citenamefont {Shahtahmassebi}, \citenamefont {Roknabad},\ and\ \citenamefont
		{Gulseren}}]{fadaie2017investigation}%
	\BibitemOpen
	\bibfield  {author} {\bibinfo {author} {\bibfnamefont {M.}~\bibnamefont
			{Fadaie}}, \bibinfo {author} {\bibfnamefont {N.}~\bibnamefont
			{Shahtahmassebi}}, \bibinfo {author} {\bibfnamefont {M.R.}\ \bibnamefont
			{Roknabad}}, \ and\ \bibinfo {author} {\bibfnamefont {O.}~\bibnamefont
			{Gulseren}},\ }\bibfield  {title} {\enquote {\bibinfo {title} {Investigation
				of new two-dimensional materials derived from stanene},}\ }\href {\doibase
		10.1016/j.commatsci.2017.05.041} {\bibfield  {journal} {\bibinfo  {journal}
			{Comp. Mater. Sci}\ }\textbf {\bibinfo {volume} {137}},\ \bibinfo {pages}
		{208--214} (\bibinfo {year} {2017})}\BibitemShut {NoStop}%
	\bibitem [{\citenamefont {Houssa}\ \emph {et~al.}(2016)\citenamefont {Houssa},
		\citenamefont {{van den Broek}}, \citenamefont {Iordanidou}, \citenamefont
		{Lu}, \citenamefont {Pourtois}, \citenamefont {Locquet}, \citenamefont
		{Afanas'ev},\ and\ \citenamefont {Stesmans}}]{ttt}%
	\BibitemOpen
	\bibfield  {author} {\bibinfo {author} {\bibfnamefont {Michel}\ \bibnamefont
			{Houssa}}, \bibinfo {author} {\bibfnamefont {Bas}\ \bibnamefont {{van den
					Broek}}}, \bibinfo {author} {\bibfnamefont {Konstantina}\ \bibnamefont
			{Iordanidou}}, \bibinfo {author} {\bibfnamefont {Anh Khoa~Augustin}\
			\bibnamefont {Lu}}, \bibinfo {author} {\bibfnamefont {Geoffrey}\ \bibnamefont
			{Pourtois}}, \bibinfo {author} {\bibfnamefont {Jean-Pierre}\ \bibnamefont
			{Locquet}}, \bibinfo {author} {\bibfnamefont {Valery}\ \bibnamefont
			{Afanas'ev}}, \ and\ \bibinfo {author} {\bibfnamefont {Andr{\'e}}\
			\bibnamefont {Stesmans}},\ }\bibfield  {title} {\enquote {\bibinfo {title}
			{Topological to trivial insulating phase transition in stanene},}\ }\href
	{\doibase 10.1007/s12274-015-0956-y} {\bibfield  {journal} {\bibinfo
			{journal} {Nano Res.}\ }\textbf {\bibinfo {volume} {9}},\ \bibinfo {pages}
		{774--778} (\bibinfo {year} {2016})}\BibitemShut {NoStop}%
	\bibitem [{\citenamefont {Fadaie}\ \emph {et~al.}(2016)\citenamefont {Fadaie},
		\citenamefont {Shahtahmassebi},\ and\ \citenamefont
		{Roknabad}}]{fadaie2016effect}%
	\BibitemOpen
	\bibfield  {author} {\bibinfo {author} {\bibfnamefont {M.}~\bibnamefont
			{Fadaie}}, \bibinfo {author} {\bibfnamefont {N.}~\bibnamefont
			{Shahtahmassebi}}, \ and\ \bibinfo {author} {\bibfnamefont {M.~R.}\
			\bibnamefont {Roknabad}},\ }\bibfield  {title} {\enquote {\bibinfo {title}
			{Effect of external electric field on the electronic structure and optical
				properties of stanene},}\ }\href {\doibase 10.1007/s11082-016-0709-5}
	{\bibfield  {journal} {\bibinfo  {journal} {Opt. Quant. Electron}\ }\textbf
		{\bibinfo {volume} {48}} (\bibinfo {year} {2016}),\
		10.1007/s11082-016-0709-5}\BibitemShut {NoStop}%
	\bibitem [{\citenamefont {Moore}\ and\ \citenamefont
		{Balents}(2007)}]{moore2007topological}%
	\BibitemOpen
	\bibfield  {author} {\bibinfo {author} {\bibfnamefont {J.~E.}\ \bibnamefont
			{Moore}}\ and\ \bibinfo {author} {\bibfnamefont {L.}~\bibnamefont
			{Balents}},\ }\bibfield  {title} {\enquote {\bibinfo {title} {Topological
				invariants of time-reversal-invariant band structures},}\ }\href {\doibase
		10.1103/PhysRevB.75.121306} {\bibfield  {journal} {\bibinfo  {journal} {Phys.
				Rev. B}\ }\textbf {\bibinfo {volume} {75}},\ \bibinfo {pages} {121306}
		(\bibinfo {year} {2007})}\BibitemShut {NoStop}%
	\bibitem [{\citenamefont {Zhu}\ \emph {et~al.}(2015)\citenamefont {Zhu},
		\citenamefont {Chen}, \citenamefont {Xu}, \citenamefont {Gao}, \citenamefont
		{Guan}, \citenamefont {Liu}, \citenamefont {Qian}, \citenamefont {Zhang},\
		and\ \citenamefont {Jia}}]{ego}%
	\BibitemOpen
	\bibfield  {author} {\bibinfo {author} {\bibfnamefont {Feng-feng}\
			\bibnamefont {Zhu}}, \bibinfo {author} {\bibfnamefont {Wei-jiong}\
			\bibnamefont {Chen}}, \bibinfo {author} {\bibfnamefont {Yong}\ \bibnamefont
			{Xu}}, \bibinfo {author} {\bibfnamefont {Chun-lei}\ \bibnamefont {Gao}},
		\bibinfo {author} {\bibfnamefont {Dan-dan}\ \bibnamefont {Guan}}, \bibinfo
		{author} {\bibfnamefont {Can-hua}\ \bibnamefont {Liu}}, \bibinfo {author}
		{\bibfnamefont {Dong}\ \bibnamefont {Qian}}, \bibinfo {author} {\bibfnamefont
			{Shou-Cheng}\ \bibnamefont {Zhang}}, \ and\ \bibinfo {author} {\bibfnamefont
			{Jin-feng}\ \bibnamefont {Jia}},\ }\bibfield  {title} {\enquote {\bibinfo
			{title} {Epitaxial growth of two-dimensional stanene},}\ }\href {\doibase
		10.1038/nmat4384} {\bibfield  {journal} {\bibinfo  {journal} {Nat. Mater.}\
		}\textbf {\bibinfo {volume} {14}},\ \bibinfo {pages} {1020--1025} (\bibinfo
		{year} {2015})}\BibitemShut {NoStop}%
	\bibitem [{\citenamefont {Xu}\ \emph {et~al.}(2014)\citenamefont {Xu},
		\citenamefont {Gan},\ and\ \citenamefont {Zhang}}]{etp}%
	\BibitemOpen
	\bibfield  {author} {\bibinfo {author} {\bibfnamefont {Yong}\ \bibnamefont
			{Xu}}, \bibinfo {author} {\bibfnamefont {Zhongxue}\ \bibnamefont {Gan}}, \
		and\ \bibinfo {author} {\bibfnamefont {Shou-Cheng}\ \bibnamefont {Zhang}},\
	}\bibfield  {title} {\enquote {\bibinfo {title} {Enhanced {{Thermoelectric
						Performance}} and {{Anomalous Seebeck Effects}} in {{Topological
						Insulators}}},}\ }\href {\doibase 10.1103/PhysRevLett.112.226801} {\bibfield
		{journal} {\bibinfo  {journal} {Phys. Rev. Lett.}\ }\textbf {\bibinfo
			{volume} {112}},\ \bibinfo {pages} {226801} (\bibinfo {year}
		{2014})}\BibitemShut {NoStop}%
	\bibitem [{\citenamefont {Molignini}\ \emph {et~al.}(2017)\citenamefont
		{Molignini}, \citenamefont {{van Nieuwenburg}},\ and\ \citenamefont
		{Chitra}}]{fmf}%
	\BibitemOpen
	\bibfield  {author} {\bibinfo {author} {\bibfnamefont {Paolo}\ \bibnamefont
			{Molignini}}, \bibinfo {author} {\bibfnamefont {Evert}\ \bibnamefont {{van
					Nieuwenburg}}}, \ and\ \bibinfo {author} {\bibfnamefont {R.}~\bibnamefont
			{Chitra}},\ }\bibfield  {title} {\enquote {\bibinfo {title} {Sensing
				{{Floquet}}-{{Majorana}} fermions via heat transfer},}\ }\href {\doibase
		10.1103/PhysRevB.96.125144} {\bibfield  {journal} {\bibinfo  {journal} {Phys.
				Rev. B}\ }\textbf {\bibinfo {volume} {96}},\ \bibinfo {pages} {125144}
		(\bibinfo {year} {2017})}\BibitemShut {NoStop}%
	\bibitem [{\citenamefont {Quelle}\ \emph {et~al.}(2016)\citenamefont {Quelle},
		\citenamefont {Cobanera},\ and\ \citenamefont {Smith}}]{morais1}%
	\BibitemOpen
	\bibfield  {author} {\bibinfo {author} {\bibfnamefont {A.}~\bibnamefont
			{Quelle}}, \bibinfo {author} {\bibfnamefont {E.}~\bibnamefont {Cobanera}}, \
		and\ \bibinfo {author} {\bibfnamefont {C.~Morais}\ \bibnamefont {Smith}},\
	}\bibfield  {title} {\enquote {\bibinfo {title} {Thermodynamic signatures of
				edge states in topological insulators},}\ }\href {\doibase
		10.1103/PhysRevB.94.075133} {\bibfield  {journal} {\bibinfo  {journal} {Phys.
				Rev. B}\ }\textbf {\bibinfo {volume} {94}},\ \bibinfo {pages} {075133}
		(\bibinfo {year} {2016})}\BibitemShut {NoStop}%
	\bibitem [{\citenamefont {Kempkes}\ \emph {et~al.}(2016)\citenamefont
		{Kempkes}, \citenamefont {Quelle},\ and\ \citenamefont {Smith}}]{morais2}%
	\BibitemOpen
	\bibfield  {author} {\bibinfo {author} {\bibfnamefont {S.~N.}\ \bibnamefont
			{Kempkes}}, \bibinfo {author} {\bibfnamefont {A.}~\bibnamefont {Quelle}}, \
		and\ \bibinfo {author} {\bibfnamefont {C.~Morais}\ \bibnamefont {Smith}},\
	}\bibfield  {title} {\enquote {\bibinfo {title} {Universalities of
				thermodynamic signatures in topological phases},}\ }\href {\doibase
		10.1038/srep38530} {\bibfield  {journal} {\bibinfo  {journal} {Scientific
				Reports}\ }\textbf {\bibinfo {volume} {6}} (\bibinfo {year} {2016}),\
		10.1038/srep38530}\BibitemShut {NoStop}%
	\bibitem [{\citenamefont {Viyuela}\ \emph
		{et~al.}(2014{\natexlab{a}})\citenamefont {Viyuela}, \citenamefont {Rivas},\
		and\ \citenamefont {Martin-Delgado}}]{delgado1}%
	\BibitemOpen
	\bibfield  {author} {\bibinfo {author} {\bibfnamefont {O.}~\bibnamefont
			{Viyuela}}, \bibinfo {author} {\bibfnamefont {A.}~\bibnamefont {Rivas}}, \
		and\ \bibinfo {author} {\bibfnamefont {M.~A.}\ \bibnamefont
			{Martin-Delgado}},\ }\bibfield  {title} {\enquote {\bibinfo {title}
			{Two-dimensional density-matrix topological fermionic phases: Topological
				uhlmann numbers},}\ }\href {\doibase 10.1103/PhysRevLett.113.076408}
	{\bibfield  {journal} {\bibinfo  {journal} {Phys. Rev. Lett.}\ }\textbf
		{\bibinfo {volume} {113}},\ \bibinfo {pages} {076408} (\bibinfo {year}
		{2014}{\natexlab{a}})}\BibitemShut {NoStop}%
	\bibitem [{\citenamefont {Viyuela}\ \emph
		{et~al.}(2014{\natexlab{b}})\citenamefont {Viyuela}, \citenamefont {Rivas},\
		and\ \citenamefont {Martin-Delgado}}]{delgado2}%
	\BibitemOpen
	\bibfield  {author} {\bibinfo {author} {\bibfnamefont {O.}~\bibnamefont
			{Viyuela}}, \bibinfo {author} {\bibfnamefont {A.}~\bibnamefont {Rivas}}, \
		and\ \bibinfo {author} {\bibfnamefont {M.~A.}\ \bibnamefont
			{Martin-Delgado}},\ }\bibfield  {title} {\enquote {\bibinfo {title} {Uhlmann
				phase as a topological measure for one-dimensional fermion systems},}\ }\href
	{\doibase 10.1103/PhysRevLett.112.130401} {\bibfield  {journal} {\bibinfo
			{journal} {Phys. Rev. Lett.}\ }\textbf {\bibinfo {volume} {112}},\ \bibinfo
		{pages} {130401} (\bibinfo {year} {2014}{\natexlab{b}})}\BibitemShut
	{NoStop}%
	\bibitem [{\citenamefont {Ezawa}(2015)}]{ezawa_monolayer_2015}%
	\BibitemOpen
	\bibfield  {author} {\bibinfo {author} {\bibfnamefont {Motohiko}\
			\bibnamefont {Ezawa}},\ }\bibfield  {title} {\enquote {\bibinfo {title}
			{Monolayer {Topological} {Insulators}: {Silicene}, {Germanene}, and
				{Stanene}},}\ }\href {\doibase 10.7566/JPSJ.84.121003} {\bibfield  {journal}
		{\bibinfo  {journal} {Journal of the Physical Society of Japan}\ }\textbf
		{\bibinfo {volume} {84}},\ \bibinfo {pages} {121003} (\bibinfo {year}
		{2015})}\BibitemShut {NoStop}%
	\bibitem [{\citenamefont {Santos}\ \emph {et~al.}(2014)\citenamefont {Santos},
		\citenamefont {Maluf},\ and\ \citenamefont {Almeida}}]{noncomm}%
	\BibitemOpen
	\bibfield  {author} {\bibinfo {author} {\bibfnamefont {Victor}\ \bibnamefont
			{Santos}}, \bibinfo {author} {\bibfnamefont {R.V.}\ \bibnamefont {Maluf}}, \
		and\ \bibinfo {author} {\bibfnamefont {C.A.S.}\ \bibnamefont {Almeida}},\
	}\bibfield  {title} {\enquote {\bibinfo {title} {Thermodynamical properties
				of graphene in noncommutative phase-space},}\ }\href {\doibase
		https://doi.org/10.1016/j.aop.2014.07.005} {\bibfield  {journal} {\bibinfo
			{journal} {Annals of Physics}\ }\textbf {\bibinfo {volume} {349}},\ \bibinfo
		{pages} {402 -- 410} (\bibinfo {year} {2014})}\BibitemShut {NoStop}%
	\bibitem [{\citenamefont {Pacheco}\ \emph {et~al.}(2003)\citenamefont
		{Pacheco}, \citenamefont {Landim},\ and\ \citenamefont
		{Almeida}}]{diracoscthermal}%
	\BibitemOpen
	\bibfield  {author} {\bibinfo {author} {\bibfnamefont {M.H}\ \bibnamefont
			{Pacheco}}, \bibinfo {author} {\bibfnamefont {R.R}\ \bibnamefont {Landim}}, \
		and\ \bibinfo {author} {\bibfnamefont {C.A.S}\ \bibnamefont {Almeida}},\
	}\bibfield  {title} {\enquote {\bibinfo {title} {One-dimensional dirac
				oscillator in a thermal bath},}\ }\href {\doibase
		https://doi.org/10.1016/S0375-9601(03)00467-5} {\bibfield  {journal}
		{\bibinfo  {journal} {Physics Letters A}\ }\textbf {\bibinfo {volume}
			{311}},\ \bibinfo {pages} {93 -- 96} (\bibinfo {year} {2003})}\BibitemShut
	{NoStop}%
	\bibitem [{\citenamefont {Boumali}(2015)}]{boumali}%
	\BibitemOpen
	\bibfield  {author} {\bibinfo {author} {\bibfnamefont {Abdelmalek}\
			\bibnamefont {Boumali}},\ }\bibfield  {title} {\enquote {\bibinfo {title}
			{Thermodynamic properties of the graphene in a magnetic field via the
				two-dimensional {{Dirac}} oscillator},}\ }\href {\doibase
		10.1088/0031-8949/90/4/045702} {\bibfield  {journal} {\bibinfo  {journal}
			{Physica Scripta}\ }\textbf {\bibinfo {volume} {90}},\ \bibinfo {pages}
		{045702} (\bibinfo {year} {2015})}\BibitemShut {NoStop}%
	\bibitem [{\citenamefont {Eshghi}\ \emph {et~al.}(2017)\citenamefont {Eshghi},
		\citenamefont {Mehraban},\ and\ \citenamefont {Ahmadi~Azar}}]{eigenspectra}%
	\BibitemOpen
	\bibfield  {author} {\bibinfo {author} {\bibfnamefont {M.}~\bibnamefont
			{Eshghi}}, \bibinfo {author} {\bibfnamefont {H.}~\bibnamefont {Mehraban}}, \
		and\ \bibinfo {author} {\bibfnamefont {I.}~\bibnamefont {Ahmadi~Azar}},\
	}\bibfield  {title} {\enquote {\bibinfo {title} {Eigenspectra and
				thermodynamic quantities in graphene under the inside and outside magnetic
				fields},}\ }\href {\doibase 10.1140/epjp/i2017-11728-9} {\bibfield  {journal}
		{\bibinfo  {journal} {The European Physical Journal Plus}\ }\textbf {\bibinfo
			{volume} {132}} (\bibinfo {year} {2017}),\
		10.1140/epjp/i2017-11728-9}\BibitemShut {NoStop}%
	\bibitem [{\citenamefont {Nasir}\ \emph {et~al.}(2010)\citenamefont {Nasir},
		\citenamefont {Khan}, \citenamefont {Tahir},\ and\ \citenamefont
		{Sabeeh}}]{weaklymodgraphene}%
	\BibitemOpen
	\bibfield  {author} {\bibinfo {author} {\bibfnamefont {R}~\bibnamefont
			{Nasir}}, \bibinfo {author} {\bibfnamefont {M~A}\ \bibnamefont {Khan}},
		\bibinfo {author} {\bibfnamefont {M}~\bibnamefont {Tahir}}, \ and\ \bibinfo
		{author} {\bibfnamefont {K}~\bibnamefont {Sabeeh}},\ }\bibfield  {title}
	{\enquote {\bibinfo {title} {Thermodynamic properties of a weakly modulated
				graphene monolayer in a magnetic field},}\ }\href
	{http://stacks.iop.org/0953-8984/22/i=2/a=025503} {\bibfield  {journal}
		{\bibinfo  {journal} {Journal of Physics: Condensed Matter}\ }\textbf
		{\bibinfo {volume} {22}},\ \bibinfo {pages} {025503} (\bibinfo {year}
		{2010})}\BibitemShut {NoStop}%
	\bibitem [{\citenamefont {Berry}(2009)}]{transitionless}%
	\BibitemOpen
	\bibfield  {author} {\bibinfo {author} {\bibfnamefont {M.~V.}\ \bibnamefont
			{Berry}},\ }\bibfield  {title} {\enquote {\bibinfo {title} {Transitionless
				quantum driving},}\ }\href {\doibase 10.1088/1751-8113/42/36/365303}
	{\bibfield  {journal} {\bibinfo  {journal} {J. Phys. A: Math. Theor.}\
		}\textbf {\bibinfo {volume} {42}},\ \bibinfo {pages} {365303} (\bibinfo
		{year} {2009})}\BibitemShut {NoStop}%
	\bibitem [{\citenamefont {Zhang}\ \emph {et~al.}(2011)\citenamefont {Zhang},
		\citenamefont {Qin}, \citenamefont {Chen}, \citenamefont {He}, \citenamefont
		{Lu}, \citenamefont {Li},\ and\ \citenamefont {Wu}}]{zqh}%
	\BibitemOpen
	\bibfield  {author} {\bibinfo {author} {\bibfnamefont {Guanhua}\ \bibnamefont
			{Zhang}}, \bibinfo {author} {\bibfnamefont {Huajun}\ \bibnamefont {Qin}},
		\bibinfo {author} {\bibfnamefont {Jun}\ \bibnamefont {Chen}}, \bibinfo
		{author} {\bibfnamefont {Xiaoyue}\ \bibnamefont {He}}, \bibinfo {author}
		{\bibfnamefont {Li}~\bibnamefont {Lu}}, \bibinfo {author} {\bibfnamefont
			{Yongqing}\ \bibnamefont {Li}}, \ and\ \bibinfo {author} {\bibfnamefont
			{Kehui}\ \bibnamefont {Wu}},\ }\bibfield  {title} {\enquote {\bibinfo {title}
			{Growth of {{Topological Insulator Bi2Se3 Thin Films}} on {{SrTiO3}} with
				{{Large Tunability}} in {{Chemical Potential}}},}\ }\href {\doibase
		10.1002/adfm.201002667} {\bibfield  {journal} {\bibinfo  {journal} {Adv.
				Funct. Mater.}\ }\textbf {\bibinfo {volume} {21}},\ \bibinfo {pages}
		{2351--2355} (\bibinfo {year} {2011})}\BibitemShut {NoStop}%
	\bibitem [{\citenamefont {Song}\ \emph {et~al.}(2010)\citenamefont {Song},
		\citenamefont {Wang}, \citenamefont {Jiang}, \citenamefont {Zhang},
		\citenamefont {Chang}, \citenamefont {Wang}, \citenamefont {He},
		\citenamefont {Chen}, \citenamefont {Jia}, \citenamefont {Wang},
		\citenamefont {Fang}, \citenamefont {Dai}, \citenamefont {Xie}, \citenamefont
		{Qi}, \citenamefont {Zhang}, \citenamefont {Xue},\ and\ \citenamefont
		{Ma}}]{swz}%
	\BibitemOpen
	\bibfield  {author} {\bibinfo {author} {\bibfnamefont {Can-Li}\ \bibnamefont
			{Song}}, \bibinfo {author} {\bibfnamefont {Yi-Lin}\ \bibnamefont {Wang}},
		\bibinfo {author} {\bibfnamefont {Ye-Ping}\ \bibnamefont {Jiang}}, \bibinfo
		{author} {\bibfnamefont {Yi}~\bibnamefont {Zhang}}, \bibinfo {author}
		{\bibfnamefont {Cui-Zu}\ \bibnamefont {Chang}}, \bibinfo {author}
		{\bibfnamefont {Lili}\ \bibnamefont {Wang}}, \bibinfo {author} {\bibfnamefont
			{Ke}~\bibnamefont {He}}, \bibinfo {author} {\bibfnamefont {Xi}~\bibnamefont
			{Chen}}, \bibinfo {author} {\bibfnamefont {Jin-Feng}\ \bibnamefont {Jia}},
		\bibinfo {author} {\bibfnamefont {Yayu}\ \bibnamefont {Wang}}, \bibinfo
		{author} {\bibfnamefont {Zhong}\ \bibnamefont {Fang}}, \bibinfo {author}
		{\bibfnamefont {Xi}~\bibnamefont {Dai}}, \bibinfo {author} {\bibfnamefont
			{Xin-Cheng}\ \bibnamefont {Xie}}, \bibinfo {author} {\bibfnamefont
			{Xiao-Liang}\ \bibnamefont {Qi}}, \bibinfo {author} {\bibfnamefont
			{Shou-Cheng}\ \bibnamefont {Zhang}}, \bibinfo {author} {\bibfnamefont
			{Qi-Kun}\ \bibnamefont {Xue}}, \ and\ \bibinfo {author} {\bibfnamefont
			{Xucun}\ \bibnamefont {Ma}},\ }\bibfield  {title} {\enquote {\bibinfo {title}
			{Topological insulator {{Bi2Se3}} thin films grown on double-layer graphene
				by molecular beam epitaxy},}\ }\href {\doibase 10.1063/1.3494595} {\bibfield
		{journal} {\bibinfo  {journal} {Appl. Phys. Lett.}\ }\textbf {\bibinfo
			{volume} {97}},\ \bibinfo {pages} {143118} (\bibinfo {year}
		{2010})}\BibitemShut {NoStop}%
	\bibitem [{\citenamefont {Bansal}\ \emph {et~al.}(2011)\citenamefont {Bansal},
		\citenamefont {Kim}, \citenamefont {Edrey}, \citenamefont {Brahlek},
		\citenamefont {Horibe}, \citenamefont {Iida}, \citenamefont {Tanimura},
		\citenamefont {Li}, \citenamefont {Feng}, \citenamefont {Lee}, \citenamefont
		{Gustafsson}, \citenamefont {Andrei},\ and\ \citenamefont {Oh}}]{bke}%
	\BibitemOpen
	\bibfield  {author} {\bibinfo {author} {\bibfnamefont {Namrata}\ \bibnamefont
			{Bansal}}, \bibinfo {author} {\bibfnamefont {Yong~Seung}\ \bibnamefont
			{Kim}}, \bibinfo {author} {\bibfnamefont {Eliav}\ \bibnamefont {Edrey}},
		\bibinfo {author} {\bibfnamefont {Matthew}\ \bibnamefont {Brahlek}}, \bibinfo
		{author} {\bibfnamefont {Yoichi}\ \bibnamefont {Horibe}}, \bibinfo {author}
		{\bibfnamefont {Keiko}\ \bibnamefont {Iida}}, \bibinfo {author}
		{\bibfnamefont {Makoto}\ \bibnamefont {Tanimura}}, \bibinfo {author}
		{\bibfnamefont {Guo-Hong}\ \bibnamefont {Li}}, \bibinfo {author}
		{\bibfnamefont {Tian}\ \bibnamefont {Feng}}, \bibinfo {author} {\bibfnamefont
			{Hang-Dong}\ \bibnamefont {Lee}}, \bibinfo {author} {\bibfnamefont {Torgny}\
			\bibnamefont {Gustafsson}}, \bibinfo {author} {\bibfnamefont {Eva}\
			\bibnamefont {Andrei}}, \ and\ \bibinfo {author} {\bibfnamefont {Seongshik}\
			\bibnamefont {Oh}},\ }\bibfield  {title} {\enquote {\bibinfo {title}
			{Epitaxial growth of topological insulator {{Bi2Se3}} film on {{Si}}(111)
				with atomically sharp interface},}\ }\href {\doibase
		10.1016/j.tsf.2011.07.033} {\bibfield  {journal} {\bibinfo  {journal} {Thin
				Solid Films}\ }\textbf {\bibinfo {volume} {520}},\ \bibinfo {pages}
		{224--229} (\bibinfo {year} {2011})}\BibitemShut {NoStop}%
	\bibitem [{\citenamefont {Chen}\ \emph {et~al.}(2011)\citenamefont {Chen},
		\citenamefont {Ma}, \citenamefont {He}, \citenamefont {Jia},\ and\
		\citenamefont {Xue}}]{cmj}%
	\BibitemOpen
	\bibfield  {author} {\bibinfo {author} {\bibfnamefont {Xi}~\bibnamefont
			{Chen}}, \bibinfo {author} {\bibfnamefont {Xu-Cun}\ \bibnamefont {Ma}},
		\bibinfo {author} {\bibfnamefont {Ke}~\bibnamefont {He}}, \bibinfo {author}
		{\bibfnamefont {Jin-Feng}\ \bibnamefont {Jia}}, \ and\ \bibinfo {author}
		{\bibfnamefont {Qi-Kun}\ \bibnamefont {Xue}},\ }\bibfield  {title} {\enquote
		{\bibinfo {title} {Molecular {{Beam Epitaxial Growth}} of {{Topological
						Insulators}}},}\ }\href {\doibase 10.1002/adma.201003855} {\bibfield
		{journal} {\bibinfo  {journal} {Adv. Mater.}\ }\textbf {\bibinfo {volume}
			{23}},\ \bibinfo {pages} {1162--1165} (\bibinfo {year} {2011})}\BibitemShut
	{NoStop}%
	\bibitem [{\citenamefont {Klaers}\ \emph {et~al.}(2017)\citenamefont {Klaers},
		\citenamefont {Faelt}, \citenamefont {Imamoglu},\ and\ \citenamefont
		{Togan}}]{Klaers}%
	\BibitemOpen
	\bibfield  {author} {\bibinfo {author} {\bibfnamefont {Jan}\ \bibnamefont
			{Klaers}}, \bibinfo {author} {\bibfnamefont {Stefan}\ \bibnamefont {Faelt}},
		\bibinfo {author} {\bibfnamefont {Atac}\ \bibnamefont {Imamoglu}}, \ and\
		\bibinfo {author} {\bibfnamefont {Emre}\ \bibnamefont {Togan}},\ }\bibfield
	{title} {\enquote {\bibinfo {title} {Squeezed thermal reservoirs as a
				resource for a nanomechanical engine beyond the carnot limit},}\ }\href
	{\doibase 10.1103/PhysRevX.7.031044} {\bibfield  {journal} {\bibinfo
			{journal} {Phys. Rev. X}\ }\textbf {\bibinfo {volume} {7}},\ \bibinfo {pages}
		{031044} (\bibinfo {year} {2017})}\BibitemShut {NoStop}%
\end{thebibliography}

%

\end{document}